\documentclass[iop]{emulateapj}
\usepackage{rotating} 
\usepackage{color}
\usepackage{amsmath} 
\newcommand{\kms}       {\mbox{km s$^{-1}$}}%
\newcommand{\msun}      {\mbox{$M_\odot$}}%
\newcommand{\mdynhat}  {\mbox{$\hat{M}_{halo}^{dyn}$}}%
\newcommand{\mdynhybrid}  {\mbox{$M_{halo}^{H-dyn}$}}%

\shortauthors{Eckert et al.}
\shorttitle{Baryonic Collapse Efficiency of Galaxy Groups}
\slugcomment{draft}

\begin{document}

\title{The Baryonic Collapse Efficiency of Galaxy Groups in the RESOLVE and ECO surveys}

\author{Kathleen D.\ Eckert\altaffilmark{1,2}, 
  Sheila J.\ Kannappan\altaffilmark{1}, 
  Claudia del P.\ Lagos\altaffilmark{3}, 
  Ashley D.\ Baker\altaffilmark{1,2}, 
  Andreas A.\ Berlind\altaffilmark{4},
  David V.\ Stark\altaffilmark{5},
  Amanda J.\ Moffett\altaffilmark{4},
  Zachary Nasipak\altaffilmark{1}, and
  Mark A.\ Norris\altaffilmark{6}
  }

\altaffiltext{1}{Department of Physics and Astronomy, University of
  North Carolina, 141 Chapman Hall CB 3255, Chapel Hill, NC 27599,
  USA; keckert@physics.unc.edu} 

\altaffiltext{2}{Department of Physics and Astronomy, University of
  Pennsylvania, 209 South 33rd St, Philadelphia, PA 19104-6396, USA}

\altaffiltext{3}{International Centre
  for Radio Astronomy Research (ICRAR), The University of Western
  Australia, 35 Stirling Highway, Crawley, WA 6009, Australia}

\altaffiltext{4}{Department of Physics and Astronomy, Vanderbilt
  University, PMB 401807, 2401 Vanderbilt Place, Nashville, TN
  37240-1807, USA}

\altaffiltext{5}{Kavli IPMU (WPI), UTIAS, The University of Tokyo, Kashiwa,
  Chiba, 277-8583, Japan}

\altaffiltext{6}{Jeremiah Horrocks Institute, University of Central
  Lancashire, Preston, PR1 2HE, United Kingdom}

\begin{abstract}

We examine the $z = 0$ group-integrated stellar and cold baryonic
(stars + cold atomic gas) mass functions (group SMF and CBMF) and the
baryonic collapse efficiency (group cold baryonic to dark matter halo
mass ratio) using the RESOLVE and ECO survey galaxy group catalogs and
a \textsc{galform} semi-analytic model (SAM) mock catalog. The group
SMF and CBMF fall off more steeply at high masses and rise with a
shallower low-mass slope than the theoretical halo mass function
(HMF). The transition occurs at group-integrated cold baryonic mass
$M_{bary}^{cold}$ $\sim$ 10$^{11}$~\msun. The SAM, however, has
significantly fewer groups at the transition mass
$\sim$10$^{11}$~\msun{} and a steeper low-mass slope than the data,
suggesting that feedback is too weak in low-mass halos and conversely
too strong near the transition mass. Using literature prescriptions to
include hot halo gas and potential unobservable galaxy gas produces a
group BMF with slope similar to the HMF even below the transition
mass. Its normalization is lower by a factor of $\sim$2, in agreement
with estimates of warm-hot gas making up the remaining difference. We
compute baryonic collapse efficiency with the halo mass calculated two
ways, via halo abundance matching (HAM) and via dynamics (extended all
the way to three-galaxy groups using stacking). Using HAM, we find
that baryonic collapse efficiencies reach a flat maximum for groups
across the halo mass range of \mbox{$M_{halo}$ $\sim$
  10$^{11.4-12}$~\msun}, which we label ``nascent groups.'' Using
dynamics, however, we find greater scatter in baryonic collapse
efficiencies, likely indicating variation in group hot-to-cold baryon
ratios. Similarly, we see higher scatter in baryonic collapse
efficiencies in the SAM when using its true groups and their group
halo masses as opposed to friends-of-friends groups and HAM masses.



\end{abstract}


\section{Introduction}
\label{sec:intro}

Galaxies form and evolve within the context of their local
environment, which can be characterized by group dark matter halos. At
$z = 0$ galaxies in low-mass halos tend to be star forming with
abundant cold gas, while in large groups and clusters, the galaxy
population is quenched of star formation with little cold gas (e.g.,
\citealp{1973MNRAS.165..231D,1983AJ.....88..483K,1984ARA&A..22..445H}). In
the largest clusters, the dominant baryonic component is the hot X-ray
emitting gas (e.g.,
\citealp{1977MNRAS.181P..25M,2009ApJ...703..982G}), while in
lower-mass halos, the halo gas temperatures are too low to emit
X-rays, presumably leaving the majority of the gas in an unobservable
warm-hot state (i.e., the warm-hot intergalactic medium WHIM,
\citealp{2006ApJ...650..560C}). The collapsed baryons (in the form of
stars and cold gas) dominate the \textit{observable} baryonic
component of such low-mass groups.


%

   
     

Previous work examining the baryonic content of clusters has used
X-ray data to study the hot gas, finding that the halo gas dominates
the baryonic content for group halos with masses
$>$10$^{13-13.5}$~\msun{} and that even in the highest mass clusters
probed ($\sim$10$^{15}$~\msun), the universal baryon fraction is not
reached (e.g.,
\citealp{2004AJ....128.2022R,2007ApJ...666..147G,2009ApJ...703..982G,2011MNRAS.412..947B}). These
works use cluster member dynamics or X-ray luminosity calibrations to
measure halo masses. For lower mass groups for which X-rays are
difficult to detect and galaxy dynamics harder to measure (due to few
members), studies have used halo abundance matching (HAM) or the halo
occupation distribution (HOD) method to study the stellar content of
groups
\citep{2010ApJ...710..903M,2012ApJ...746...95L,2013ApJ...770...57B}. These
studies find that the group stellar fraction (group stellar mass
divided by group halo mass) peaks at halo masses
$\sim$10$^{12}$~\msun{} and decreases towards higher and lower halo
masses.





These previous studies have focused on the stellar content of groups,
leaving out the contribution from cold gas (the reservoir for future
star formation), which can dominate the galaxy mass for galaxies with
cold baryonic mass \mbox{$M_{bary}^{cold}$ $<$ 10$^{9.9}$ \msun}, the
gas richness threshold mass defined in \citet{2013ApJ...777...42K},
hereafter K13. Even at higher galaxy masses, star-forming galaxies
have HI gas-to-stellar mass ratios typically ranging from 0.1--1,
(e.g.,
\citealp{2013MNRAS.436...34C,2013ApJ...777...42K,2015MNRAS.452.2479B}). In
this work, we define the term ``cold baryonic mass'' to mean the mass
in stars and cold atomic gas (see \S \ref{sec:grpint}), neglecting
other cold gas components. In previous work, we showed that the
low-mass slope of the galaxy cold baryonic mass function rises more
steeply than that of the stellar mass function
(\citealp{2016ApJ...824..124E}, hereafter E16). We also found complex
structure after breaking the baryonic mass function into different
group halo mass regimes. In the intermediate group halo mass regime
$\sim$10$^{11.4-12}$ \msun, we found a flat low-mass slope,
potentially a signature of group formation processes such as stripping
and merging. We refer to groups in this mass range as ``nascent
groups,'' where galaxies first start to come together to form larger
structures.

These results motivate our desire to study group-integrated mass
functions and the group baryonic collapse efficiency (the cold
baryonic group mass divided by the group halo mass). To perform this
study we use two volume-limited surveys with groups ranging in halo
mass from $\sim$10$^{11}$~\msun{} to 10$^{14.5}$~\msun. The smaller,
RESOLVE-B, is complete to an individual galaxy cold baryonic mass
limit of \mbox{$M_{bary}^{cold}$ $\sim$ 10$^{9.1}$ \msun}. The larger,
ECO, encompasses the RESOLVE-A subvolume and is complete to galaxy
\mbox{$M_{bary}^{cold}$ $\sim$ 10$^{9.4}$ \msun}. We also construct a
mock catalog from the \textsc{galform} semi-analytic model (SAM) of
\citet{2014MNRAS.439..264G} to compare with the data.


In \S \ref{sec:dataandmethods} we describe the data and methods used
in this work to measure the mass of groups in terms of stellar, cold
baryonic, and group halo mass. In \S \ref{sec:groupmfsandbfrac} we
analyze the group-integrated stellar and cold baryonic mass functions
(SMF and CBMF) and examine the stellar and cold baryonic fractions of
groups, finding a broad peak in baryonic collapse efficiency from
10$^{11.4-12}$ \msun{} across the nascent group regime. In \S
\ref{sec:discussion} we discuss the implications of our results on
nascent group formation and undetected forms of gas. Finally, in \S
\ref{sec:conclusions} we summarize our conclusions.


 







%



\section{Data And Methods}
\label{sec:dataandmethods}

Below we present a brief overview of the two data sets used in this
work, including their relative merits. We also present a description
of the data used to construct the group-integrated properties for the
two data sets. Finally, we describe the mock catalog created from the
\textsc{galform} SAM of \citet{2014MNRAS.439..264G}.

\subsection{Data Sets}
\label{sec:datasets}

In this work we use two data sets, the REsolved Spectroscopy of a
Local VolumE survey (RESOLVE, Kannappan et al.\ in prep.)  and the
Environmental COntext catalog (ECO, \citealp{2015ApJ...812...89M},
hereafter M15). Both data sets are volume-limited and have been
constructed using the SDSS main redshift survey
\citep{2002AJ....124.1810S}, filling in incompleteness due to
fiber-collisions and pipeline photometry issues (see
\citealp{2005ApJ...631..208B}) with data from several other redshift
surveys as described in E16. For both surveys we define membership
based on group redshift using a buffer region to recover galaxies
whose peculiar velocities place them outside the survey limits (see \S
\ref{sec:envandhmass}). While the RESOLVE survey has greater
completeness and deeper photometric and HI data, the ECO catalog
covers a much larger volume, providing better statistics and a wider
range of group halo masses.

The RESOLVE survey covers a $>$50,000~Mpc$^3$ volume over two
equatorial strips ranging in redshift from 4500--7000~\kms{} (see
\citealp{2015ApJ...810..166E}, hereafter E15, for more detail). The
$\sim$13,700~Mpc$^3$ RESOLVE-B footprint coincides with SDSS Stripe
82, while the larger RESOLVE-A is surrounded by the ECO
catalog. RESOLVE-B has extra redshift completeness due to repeated
observations by the SDSS (see E16).  Due to the extra redshift
completeness, we have dropped the RESOLVE-B luminosity completeness
limit to \mbox{$M_{r,tot}$ = $-17.0$}, below the nominal luminosity
completeness limit of \mbox{$M_{r,tot}$ = $-17.33$}, which corresponds
to the SDSS apparent magnitude survey limit of 17.77 at the outer
redshift boundary using the RESOLVE total magnitudes from
\citet{2015ApJ...810..166E}. The RESOLVE-B volume contains 486
galaxies brighter than this limit and 344 groups, 286 of which have
\mbox{$N = 1$} member.


RESOLVE-B is covered by deep $ugriz$ coadds in the SDSS
\citep{2011ApJS..193...29A}, as well as shallow $JHK$ 2MASS
\citep{2006AJ....131.1163S} and deeper $YHK$ UKIDSS data
\citep{2008MNRAS.384..637H}. In addition it has nearly complete
coverage by the \textit{GALEX} MIS depth survey ($\sim$ 1500s) in the
NUV \citep{2007ApJS..173..682M}, plus \textit{Swift} uvm2 imaging for
19 galaxies (E15). The RESOLVE HI survey, presented in
\citet{2016ApJ...832..126S}, provides unconfused (or deconfused) HI
detections or strong upper limits (\mbox{1.4$M_{HI}$ $<$
  0.05$M_{star}$}) for 87\% of galaxies brighter than \mbox{$M_{r,tot}$ =
  $-17.0$} or having estimated \mbox{$M_{bary}^{cold}$ $>$ 10$^{9.0}$
  \msun}, based on calibrations of the relationship between
gas-to-stellar mass ratio and galaxy color (the photometric gas
fractions technique described in E15).

The ECO catalog covers a volume of $\sim$442,700 Mpc$^{3}$, which is
$\sim$32 times larger than RESOLVE-B and encompasses RESOLVE-A. While
less complete in terms of redshift coverage, the ECO volume provides
statistical power that the smaller RESOLVE-B subvolume cannot, having
9443 galaxies brighter than the luminosity limit of $-17.33$ and 6746
groups of which 5723 are groups of \mbox{$N = 1$}.

While ECO has uniform shallow coverage over $ugrizJHK$ from SDSS and
2MASS, deeper imaging from UKIDSS is limited to the RESOLVE-A region
and MIS depth NUV from \textit{GALEX} covers $\sim$45\% of ECO
(including most of RESOLVE-A). Fractional-mass limited HI data are
available for the RESOLVE-A subvolume within ECO, providing a similar
quality of data compared to RESOLVE-B. Additional coverage is provided
by the flux-limited 21cm ALFALFA survey's $\alpha$40 catalog
\citep{2011AJ....142..170H}, which yields HI detections for galaxies
with \mbox{$M_{HI}$ $\gtrsim$ 10$^{9}$ \msun{}} at ECO redshifts. We
have computed upper limits for galaxies with ALFALFA non-detections,
but $\sim$84\% of those are weak (i.e., \mbox{1.4$M_{HI}^{limit}$ $>$
  0.05$M_{star}$}). For ECO galaxies without HI data or having only a
weak upper limit, we rely on gas mass estimates using the photometric
gas fractions technique described in E15, which provides full
probability distributions for the gas mass, not just point estimates
as in previous work (e.g.,
\citealp{2004ApJ...611L..89K,2012MNRAS.424.1471L,2013MNRAS.436...34C}).

In E16, we computed galaxy stellar and baryonic mass completeness
limits for the RESOLVE-B and ECO volumes by examining the stellar and
baryonic mass-to-light ratio distributions near each survey's
respective luminosity completeness limit. For RESOLVE-B, we find that
the stellar and baryonic mass completeness limits are \mbox{$M_{star}$
  = 10$^{8.7}$ \msun{}} and \mbox{$M_{bary}^{cold}$ = 10$^{9.1}$
  \msun}. For ECO, they are \mbox{$M_{star}$ = 10$^{8.9}$
  \msun{}} and \mbox{$M_{bary}^{cold}$ = 10$^{9.4}$ \msun}.

To determine our group mass completeness limits, we note that at low
group mass, we are dominated by \mbox{$N = 1$} groups, so these galaxy
mass completeness limits should roughly translate to group-integrated
mass limits. There may, however, be groups consisting entirely of
galaxies below our luminosity completeness limit (such as dwarf
associations; \citealp{2006AJ....132..729T}). To quantify how many
such groups we may be missing, we examine the number of groups in
RESOLVE-B with group-integrated stellar or cold baryonic mass greater
than the shallower ECO mass completeness limits, but having no galaxy
brighter than the ECO luminosity limit ($-17.33$). We find $<$1\% of
RESOLVE-B groups fit this criteria, implying that our galaxy
completeness limits are sufficient.

Due to the superior spectroscopic completeness of RESOLVE-B relative
to SDSS (E16), we can consider RESOLVE-B to be a truly complete data
set. For ECO, however, we know that we are missing galaxies due to
both fiber collisions and surface brightness incompleteness, despite
efforts to account for galaxies through merging of several
spectroscopic surveys (M15, E16).  To address this incompleteness in
ECO, E16 computed galaxy completeness corrections as a function of
luminosity and color by comparing the completeness of RESOLVE-B and
ECO relative to the main SDSS redshift survey over luminosity-color
space (see M15 and E16 details). These completeness corrections have
been applied as weights in the galaxy mass functions in E16. We
describe how we translate the weights to group completeness
corrections in \S \ref{sec:ecocc}.

\subsection{Photometry \& Galaxy Stellar and Cold Baryonic Masses}
\label{sec:photandmass}

In this work we use reprocessed photometry, as described in E15 and
M15 for the RESOLVE and ECO data sets respectively. Our reprocessing
addresses several issues in the catalog photometry. For the SDSS data,
we use the improved sky background subtraction of
\citet{2011AJ....142...31B}, and for the IR data we perform additional
custom background subtraction. By enforcing the same elliptical
apertures (based on the high S/N $gri$ coadded image) across all
bands, we are able to measure total galaxy magnitudes in all bands
using three non-parametric methods comparison of which yields
systematic error estimates. Our methods allow for color gradients in
galaxies as opposed to suppressing the algorithms used for the SDSS
catalog photometry do \citep{2002AJ....123..485S}.

These improvements yield brighter magnitudes, larger radii, overall
bluer colors, and more real scatter in color (see Figures 3 and 4 of
E15). These last two points imply that galaxy star formation rates are higher
and star formation histories are more varied than previously
reported. In K13, which used similarly processed photometry for the
Nearby Field Galaxy Survey (NFGS, \citealp{2001Ap&SS.276.1151J}),
low-mass gas-rich galaxies, traditionally regarded as poor star
formers, were found to be doubling their masses over the last Gyr.

RESOLVE and ECO stellar masses (E15, M15) were computed using the
Bayesian spectral energy distribution (SED) fitting approach described
in K13 (see also \citealp{2007ApJ...657L...5K}). The code produces a
likelihood weighted mass distribution for each galaxy based on the
full model grid considered. Briefly, the grid consists of an old and
young stellar population, each populated with a Chabrier IMF. The old
stellar population is modeled as a burst with age ranging from 2--12
Gyr. The young stellar population is modeled either as continuous star
formation starting 1015 Myr ago and continuing to a turnoff sometime
in the last 0--195 Myr or as a single quenching burst with age ranging
from 360--1015 Myr. The young stellar population can contribute from
0.1\%--94.1\% of the stellar mass. The grid includes four
metallicities ranging from \mbox{$Z$ = 0.004--0.05}, and eleven
optical depth dust values (ranging from $\tau_{v}$ = 0--1.2) are
applied to the young stellar population using the dust law from
\citet{2001PASP..113.1449C}. We generally use the full mass likelihood
distribution in this work, but when we assign a single value for the
galaxy's stellar mass, we take the median of the likelihood weighted
stellar mass distribution.



As previously mentioned, cold baryonic mass in this work is defined as
the stars plus cold atomic gas mass. Generally the atomic gas
dominates the cold gas mass of galaxies, although large spirals may
have significant reservoirs of molecular gas. The total gas mass in
large spirals, however, is typically $\lesssim$ half the stellar mass
\citep{1998A&A...331..451C,2013ApJ...777...42K,2014A&A...564A..66B}. RESOLVE-B
and ECO both have HI data available with varying depth and
coverage. While RESOLVE's coverage is fractional mass limited and
nearly complete, ECO has fractional mass limited data only in the
RESOLVE-A subvolume and relies on the flux-limited ALFALFA survey
elsewhere, which provides mostly weak upper limits. In this work, we
define the atomic gas as 1.4$M_{HI}$ to account for the contribution
from helium.

To supplement the HI data, we use the photometric gas fraction (PGF)
technique to estimate gas-to-stellar mass (G/S) ratios as described in
E15. The estimators are based on a model fit to the 2D distribution of
log(G/S) vs.\ color (or ``modified color,'' a linear combination of
color and axial ratio) to produce log(G/S) distributions for each
galaxy. These estimates of log(G/S) are created using the RESOLVE-A
data set and are therefore ideal for use on volume-limited surveys,
as validated by testing on the RESOLVE-B HI data set in E15. 


To compute cold baryonic mass, we perform a ``pseudo-convolution'' of
the stellar mass likelihood distribution for a given galaxy with the
HI mass likelihood distribution implied by its HI data (for good
detections) or inferred from its PGF-estimated log(G/S) distribution
(for missing, low S/N, or badly confused detections). The details are
provided in E16. This algorithm results in a cold baryonic mass
likelihood distribution, from which we can take the median if a single
value for the galaxy's baryonic is necessary. (We use the full
distribution by default.)

\subsection{Group Stellar, Cold Baryonic, and Halo Masses}
\label{sec:envandhmass}

The fact that RESOLVE and ECO are volume limited enables optimal group
finding, for which we use the Friends-of-Friends (FOF) algorithm from
\citet{2006ApJS..167....1B}. This algorithm links galaxies that are
within a specified projected and line-of-sight linking length into
groups. The projected and line-of-sight linking lengths determined in
\citet{2006ApJS..167....1B}, respectively \mbox{$b_{\perp}$ = 0.14}
and \mbox{$b_{\parallel}$ = 0.75} times the mean separation between
objects, were designed to reproduce the multiplicity function and
projected sizes of groups with \mbox{$N > 10$} members. Based on our
own work as well as that of \citet{2014MNRAS.440.1763D} and
\citet{2011MNRAS.416.2640R}, we use projected and line-of-sight
linking lengths better geared toward recovery of low-$N$ groups and
dynamical masses: \mbox{$b_{\perp}$ = 0.07} and \mbox{$b_{\parallel}$
  = 1.1} times the mean separation between objects (for more detail
see \S 3.5.1 of E16). Since RESOLVE-B is a small volume (and overdense
due to cosmic variance, E16), we fix its linking lengths to equal
those computed for a version of ECO that extends to \mbox{$M_{r,tot}$
  = $-17.0$}, i.e., a version of ECO with depth analogous to RESOLVE-B
but without its overdensity (see M15). After running the FOF code,
each galaxy is assigned to a group. We consider galaxies that are
identified as being alone in their halo as \mbox{$N = 1$} groups with
isolated ``central'' galaxies.
 
As a consequence of the FOF group finding algorithm, many isolated $N
= 1$ groups are falsely linked into pairs. To cut down the number of
false pairs, we use a mock catalog, for which we know the true pairs,
to identify a region in $\Delta$cz -- $R_{proj}$ space containing 95\%
of true pairs. Breaking up all pairs outside this region into groups
of $N = 1$, the percentage of true pairs in the FOF group catalog
increases from 62\% to 73\%. For further information on the algorithm
used to break up false pairs, see the Appendix.

To describe the mass content of groups in this work, we use three
different types of metrics: the group-integrated stellar and cold
baryonic mass (group $M_{star}$ and $M_{bary}^{cold}$, the group halo
mass determined through halo abundance matching ($M_{halo}^{HAM}$),
and the group total mass determined from dynamics
($M_{halo}^{dyn}$). We have also provided machine readable tables with
the group information and quantities described in the following
sections for the RESOLVE and ECO galaxy catalogs. The columns of the
data provided in the two tables are given in Table
\ref{tb:groupinfotable}\footnote{The coordinates provided in the table
  have been updated to reflect changes described in the erratum to E15
  that affect 29 galaxies in RESOLVE-B. These updated coordinates
  change the measured $R_{proj}$ for affected groups by less than
  2\%. Group stellar masses have not been updated to reflect the
  changes described in the erratum as most differences in the stellar
  mass estimates are $<$ 0.03 dex}.

\begin{deluxetable}{ll}
\tablecaption{RESOLVE and ECO Group Catalog Description}
\tablehead{\colhead{Column} & \colhead{Description}}
\startdata
1 & RESOLVE or ECO galaxy ID \\
2 & group ID \\
3 & group N \\
4 & group RA \\
5 & group Dec \\
6 & group cz \\
7 & HAM halo mass ($M_{halo}^{HAM}$, based on group $L_r$) \\
8 & HAM halo mass ($M_{halo}^{HAM}$, based on group M$_{star}$) \\
9 & dynamical halo mass ($\hat{M}_{halo}^{dyn}$, scaled by A = 9.9) \\
10 & stacked dynamical halo mass (scaled by A = 9.9) \\
11 & hybrid dynamical halo mass ($M_{halo}^{H-dyn}$) \\
12 & group-integrated stellar mass (group $M_{star}$) \\
13 & group-integrated cold baryonic mass (group $M_{bary}^{cold}$) \\
14 & central galaxy flag (brightest galaxy in $L_r$) \\
15 & group velocity dispersion ($\sigma_{grp}$, using Gapper method) \\
16 & group projected radius ($R_{proj}$, using percentile method) \\
\enddata
\label{tb:groupinfotable}
\end{deluxetable}

\subsubsection{Group-Integrated Stellar and Cold Baryonic Mass}
\label{sec:grpint}

The group-integrated stellar and cold baryonic masses (group
$M_{star}$ and $M_{bary}^{cold}$) are the respective sums of the
stellar and cold baryonic masses of all galaxies within the group. To
compute the likelihood distributions of these integrated masses for
each group, we use a pseudo-convolution method similar to the method
used to compute galaxy baryonic mass in E16. We do this so that we can
use the mass likelihood distributions for each group with the
cross-bin sampling technique of E16 to determine smooth
group-integrated mass functions and uncertainty bands (see \S
\ref{sec:groupmfs}).

Briefly, for groups with \mbox{$N = 1$} member, the group-integrated
mass likelihood distribution is the mass likelihood distribution of
the single galaxy. For groups with \mbox{$N > 1$} members, we start
with the two least massive galaxies. First, we compute the mass
likelihood distributions of both galaxies divided into bins of linear
spacing $\Delta$M, which can range from 1/100 to 1/2 of the smallest
mass with likelihood $>$1e-4. The range accounts for the potentially
large difference in mass between the two galaxies (possibly a factor
of 10--100) to keep the calculation from taking too long. We then
perform the ``pseudo-convolution'' by computing the new mass and
likelihood for each possible mass combination of the two
galaxies. This pseudo-convolution is repeated with each resulting mass
likelihood distribution and successively more massive galaxy in the
group (updating $\Delta$M for each round) to produce the group mass
likelihood distribution. To assign a specific stellar or baryonic mass
value to a group, we use the median of its mass likelihood
distribution. The typical uncertainties on these group-integrated
masses are comparable to the factor of $\sim$1.5--2 uncertainties on
stellar masses from SED fitting.


We can estimate the neglected stellar and cold baryonic mass
contribution from satellites below the survey limit floor by using the
satellite mass functions from E16. First we normalize the satellite
mass functions in each group halo mass regime presented in E16
(divisions at 10$^{11.4}$~\msun, 10$^{12}$~\msun, and
10$^{13.5}$~\msun). Then we fit a line to the low-mass slope to
extrapolate the satellite mass function below our survey mass limits,
and we integrate the extrapolated total mass in satellites from our
survey limit down to $M_{star}$ or \mbox{$M_{bary}^{cold}$ = 10$^{6}$
  \msun}. In the nascent group halo mass regime \mbox{($M_{halo}$ =
  10$^{11.4-12}$ \msun)}, we find that satellites below our mass
limits contribute an additional $\sim$2\%--4\% to group $M_{star}$ and
$\sim$7\%--8\% to group $M_{bary}^{cold}$. In the large group halo mass
and cluster mass regimes (\mbox{$M_{halo}$ = 10$^{12.0-13.5}$ \msun{}}
and \mbox{$M_{halo}$ $>$ 10$^{13.5}$ \msun}) the contribution falls to
$\sim$1.5\% for group $M_{star}$ and $\sim$3\%--4\% for group
$M_{bary}^{cold}$. To estimate the contribution from satellites below
our luminosity limit for our lowest group halo mass regime (for which
the satellite mass function is mostly incomplete), we scale the
extrapolated slopes for the nascent and large-group halo
satellite mass functions to estimate a range. In this halo mass
regime, the satellite contribution to group $M_{star}$ ranges from
3\%--8\% and to group $M_{bary}^{cold}$ ranges from 8\%--14\%, depending on
the slope used.

While these contributions from galaxies below our survey floor have
not been added to our group-integrated stellar and cold baryonic
masses, we do include mass from missing galaxies above our mass limit
in ECO using the group completeness corrections described in \S
\ref{sec:ecocc}.

\subsubsection{Group Halo Abundance Matching}
\label{sec:ham}
 
Group halo abundance matching (HAM) uses the cumulative number density
of groups based on some group quantity (such as group luminosity) and
matches the groups to halos of corresponding cumulative number density
in simulations. In this work we perform HAM using the halo mass
function (HMF) from \citet{2006ApJ...646..881W}, adopting a
cosmological model with \mbox{$H_0$ = 70 \kms$Mpc^{-1}$},
\mbox{$\Omega_m$ = 0.3}, and \mbox{$\sigma_8$ = 0.9}. We use both the
group-integrated $r$-band luminosity (group $L_{r}$) down to the
survey absolute magnitude floor and group $M_{star}$ to perform the
matching. In particular, we note that galaxy $L_r$ correlates more
tightly with cold-baryonic mass than with stellar mass (K13),
suggesting that group $L_r$ should also correlate tightly with group
M$_{bary}^{cold}$. \citet{2006ApJ...646..881W} use the standard
simulation linking length parameter b = 0.2, finding that these FOF
halos are roughly equivalent to halos defined at $M_{280b}$ or at an
overdensity of 280 times the background density of matter. This method
assumes that group halo mass can be determined based on the group
properties corresponding to the stellar or cold-baryonic content of
groups and that every halo is populated by a galaxy. The first
assumption implies a monotonic relationship between the HAM group halo
mass and group $M_{star}$ or $M_{bary}^{cold}$ as seen in Figure
\ref{fg:mbvmg}a. This assumption will fail in the presence of
significant variations in hot baryon fraction.

\subsubsection{Group Dynamical Masses} 
\label{sec:dynhn}

For groups with multiple members, we compute dynamical masses using
the relative velocities and projected distances of the galaxies from
the group center. We use the virial theorem to calculate the dynamical
mass.

\begin{equation}
\hat{M}_{halo}^{dyn} = A\frac{\sigma_{grp}^{2}R_{proj}}{G}
\label{eq:mdyn}
\end{equation}

\noindent
where $\sigma_{grp}$ is the velocity dispersion of the group,
$R_{proj}$ is the projected radius of the group, and $A$ is a
multiplicative scale factor. The scale factor accounts for the
projected radius not being the virial radius.

To measure $\sigma_{grp}$, we use the Gapper method
\citep{1990AJ....100...32B}, which is more robust than a simple rms
for low-$N$ groups. The Gapper method weights the radial velocities of
the galaxies in each group using the formula:

\begin{equation}
\sigma_{Gapper} = \frac{\sqrt{\pi}}{n(n-1)}\sum_{i=1}^{N-1}\Delta v_iw_i
\label{eq:gapper}
\end{equation}

\noindent
where $N$ is the number of galaxies in the group, $\Delta v_{i}$ is
$v_{i+1} - v_{i}$ (the velocities have been ordered from smallest to
largest), and $w_{i}$ = $i(N-i)$.

The group's projected radius is measured relative to the group center
computed by taking the mean of the member galaxies' RA and Dec
coordinates. Using the technique from \citet{2011MNRAS.416.2640R}, we
order the galaxies' projected radii from the center from smallest to
largest and assign to each ordered radius the percentage of galaxies
within that radius (0\%--100\%). We then find the radius corresponding
to the 75th percentile to be the projected radius (thus the two
galaxies with percentiles bracketing 75 determine the group
radius). We use a larger percentile than the preferred 50th percentile
used in \citealp{2011MNRAS.416.2640R}, which best recovered group
radii for \mbox{$N > 20$} groups, but also developed artifacts for
low-$N$ systems due to group finding errors and the group radius
definition. These artifacts are reduced when using a larger percentile
definition, although choosing too large a percentile will result in
projected radius measurements that are susceptible to outliers.



The underlying assumption of the dynamical approach to mass estimation
is that the group halo is virialized, which may or may not be a safe
assumption. To assess the validity of this assumption, we use the
Anderson--Darling test (A--D test) following the methods of
\citet{2009ApJ...702.1199H} to look for whether the distribution of
radial velocities is consistent with a Gaussian distribution. The A--D
test is considered robust for \mbox{$N \geq 5$} systems
\citep{1986gft..book.....D}, and we find that $\sim$90\% of our groups
with $N \geq 5$ can be classified as virialized.

\begin{figure*}
\plotone{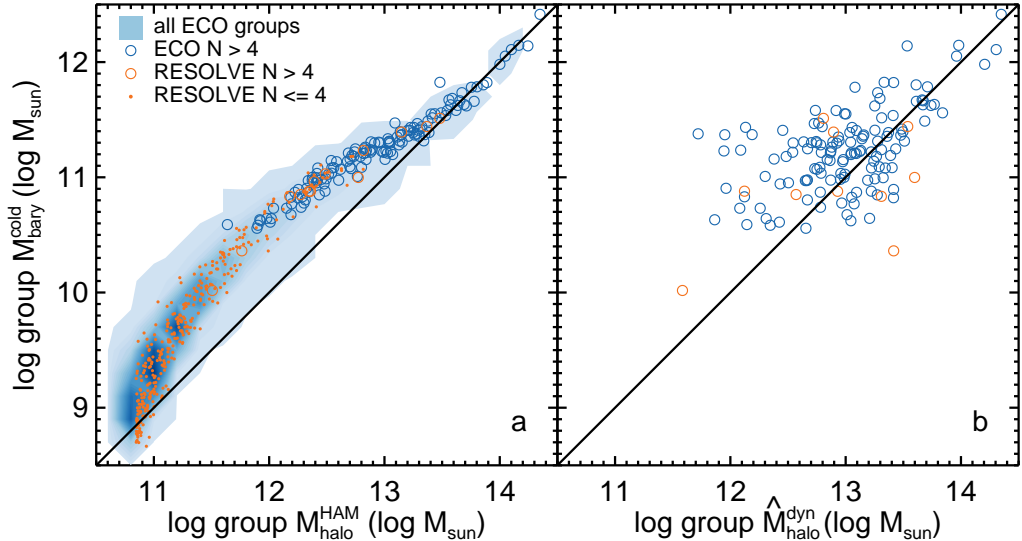}
\epsscale{1.0}
\caption{Group $M_{bary}^{cold}$ (including satellite completeness
  corrections, see \S \ref{sec:ecocc}) vs.\ $M_{halo}$ measured by (a)
  HAM and (b) dynamical estimates. ECO is shown in blue and RESOLVE is
  shown in orange. The black line shows a one-to-one relationship
  shifted down by 2 dex. Groups with \mbox{$N > 4$} are shown as open
  circles. $M_{halo}^{HAM}$ is closely correlated to the group
  $M_{bary}^{cold}$ since we have used group $r$-band luminosity
  (group $L_r$) to perform the abundance matching and $L_r$ correlates
  closely with cold baryonic mass (K13). Dynamical estimates are shown
  only for groups with \mbox{$N > 4$} members and show larger scatter
  with baryonic content at lower halo masses due to having fewer
  members for the dynamical mass calculation.}
\label{fg:mbvmg}
\end{figure*}

In Figure \ref{fg:mbvmg}, we show group $M_{bary}^{cold}$
vs.\ $M_{halo}^{HAM}$ (using group $L_r$) and vs.\ \mdynhat{} for
RESOLVE-B and ECO. The built-in relationship between group
$M_{bary}^{cold}$ and $M_{halo}^{HAM}$ is apparent, while the
relationship between group $M_{bary}^{cold}$ and \mdynhat{} shows more
scatter, especially towards lower mass groups that have fewer galaxies
with which to compute the dynamics. We have scaled \mdynhat{} by
\mbox{$A$ = 9.9}, which minimizes the offset between HAM and dynamical
group mass estimates for groups with $N > 7$. The HAM masses already
match the simulation HMF by construction.

\subsubsection{Group Dynamical Masses Through Stacking}
\label{sec:dynln}

For low-N groups, dynamical masses are less reliable and for singleton
and pair groups, they are impossible to calculate. Therefore, for
groups with $N > 2$ we stack groups of similar properties and compute
the velocity dispersion and projected radius from a larger number of
galaxies.

\begin{figure*}
\epsscale{0.75}
\plotone{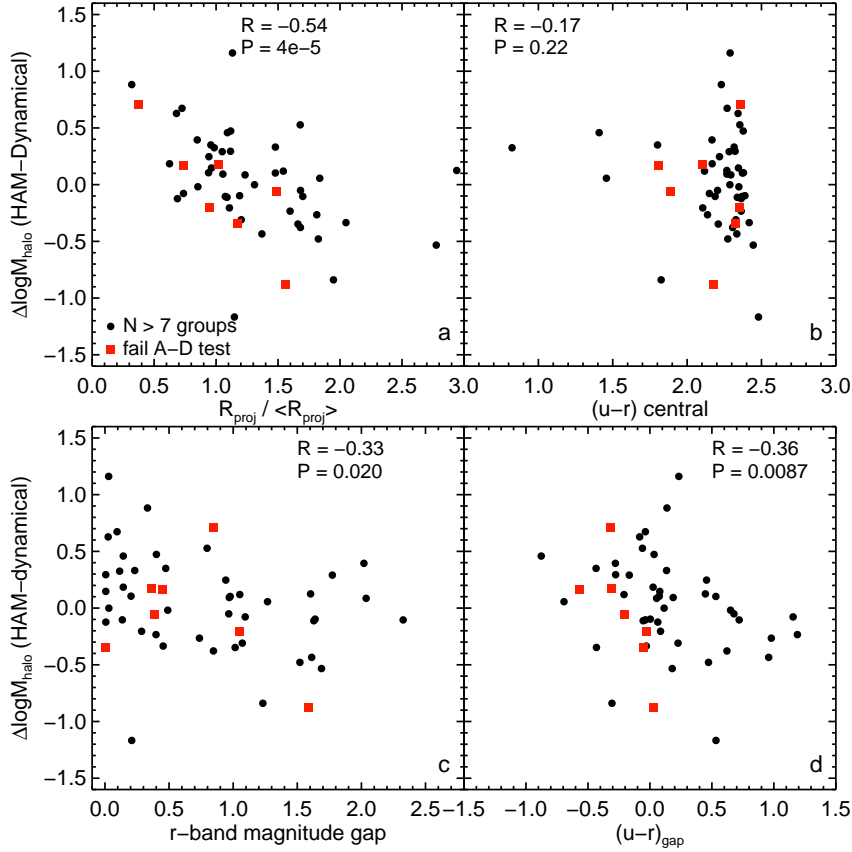}
\caption{Residual correlations between $\Delta$log$M_{halo}$ and (a)
  $R_{proj}$/$<R_{proj}>$ the projected radius normalized to the
  median projected radius at a given group $L_r$, (b) $u-r$ color of
  the central, (c) $r$-band magnitude gap, and (d) $u-r$ color
  gap. Only groups with \mbox{$N > 7$} members are shown, and
  $\Delta$log$M_{halo}$ is computed as the difference between the log
  of $M_{halo}^{HAM}$ and the log of \mdynhat{} (scaled by
  $A = 9.9$). Groups that fail the A--D test (i.e., that are not
  virialized) are shown in red. The Spearman rank correlation
  coefficient and probability of no correlation are reported as $R$
  and $P$ respectively. $R_{proj}$/$<R_{proj}>$ yields the most
  significant correlation with $\Delta$log$M_{halo}$ with the highest
  coefficient of correlation and the least scatter. The color of the
  central does not show a significant correlation with
  $\Delta$log$M_{halo}$. The $r$-band magnitude gap and $u-r$ color
  gap do show significant correlations with $\Delta$log$M_{halo}$,
  although with smaller coefficients of correlation and greater
  scatter than for $R_{proj}$/$<R_{proj}>$.}
\label{fg:resids}
\end{figure*}

To compute the stacked dynamical masses, we first need to determine
what group properties to stack on. For the first parameter we use the
group-integrated luminosity, which to first order should track the
mass of the group. For the second parameter, we have tested different
quantities that may relate to the dynamical state of the group by
examining whether they correlate with residuals from the relation
between $M_{halo}^{HAM}$ and \mdynhat{} in \mbox{$N > 7$} member
groups (Figure \ref{fg:resids}). Typically, a cutoff of $N=10$ is used
for reliable dynamical mass measurements, however we have chosen $N >
7$ because the distribution between the two variables is roughly
Gaussian in this regime and we can increase the sample of groups from
34 to 51.  The parameters are: projected radius normalized to the
median projected radius for a given group $L_r$
($R_{proj}$/$<R_{proj}>$), $u-r$ color of the central galaxy, $r$-band
magnitude gap, and $u-r$ color gap. The last two are computed between
the central (the brightest galaxy in $M_{r,tot}$) and the brightest
satellite (the second brightest galaxy in $M_{r,tot}$).

The normalized projected radius may reveal offsets in dynamical mass
to the extent that the degree of compactness relates to dynamical
status. Since $R_{proj}$ goes into the dynamical mass measurements,
there is covariance between this quantity and halo mass residuals. The
magnitude gap between the central and brightest satellite has been
used as a tool to detect groups and clusters that assembled early and
hence are more dynamically relaxed (e.g.,
\citealp{1994Natur.369..462P,2003MNRAS.343..627J}). The recent merger
history of halos, however, may enhance or diminish magnitude gaps
within groups, making them less reliable as an indicator of early
assembly \citep{2008MNRAS.386.2345V,2010MNRAS.405.1873D}. More recent
work has examined the use of galaxy color to perform age distribution
matching (along with HAM, e.g., \citealp{2013MNRAS.435.1313H}),
suggesting that the color of the central or of the entire group may be
useful for quantifying the assembly history of the group. Based on
these studies, we explore the central galaxy color as well as a
quantity that we call the color gap, which is the difference in color
between the central and brightest satellite.



We find that $R_{proj}$/$<R_{proj}>$ yields the most significant
correlation with $\Delta$log$M_{halo}$ using the Spearman rank
correlation test, with a correlation coefficient of $-0.54$ and
relatively little scatter. The $r$-band magnitude gap and $u-r$ color
gap also yield significant correlations with halo mass residuals,
although both have small correlation coefficients compared to
$R_{proj}$/$<R_{proj}>$. It is interesting that the color gap yields
such a significant correlation (albeit with large scatter), as it
suggests that deviations between dynamical and HAM mass may be related
to deviations from galaxy conformity, the empirical result showing
that satellites tend to have colors and star formation histories
similar to their central \citep{2006MNRAS.366....2W}.




We use $R_{proj}$/$<R_{proj}>$ along with group $L_r$ to stack $N > 2$
groups in bins of 0.2 and 0.25 mag respectively. We then compute
the stacked group dynamical mass for each bin, which is applied to all
groups in that bin.

\subsubsection{Final Dynamical Group Mass Estimates}
\label{sec:finaldyn}

To determine final dynamical mass estimates, we rely on a combination
of measured and stacked estimates at high $N$ and stacked and HAM
estimates at low $N$. We also calibrate the hybrid dynamical masses to
match the cumulative HMF. We label this hybrid dynamical mass
\mdynhybrid{} to distinguish it from the directly measured dynamical
masses that have been scaled by a single constant \mdynhat.

For groups with \mbox{$N \geq 15$}, we use the measured dynamical
mass. For groups with \mbox{$3 \leq N \leq 7$}, we use the stacked
dynamical mass estimates. For groups with $N$ between 7 and 15, we
transition smoothly between these two regimes by using a linear
combination of the stacked dynamical mass estimate and the directly
measured dynamical estimate as given by equation \ref{eq:lincomb}:

\begin{equation}
\mdynhybrid{} = a \times M_{halo}^{dyn} + (1-a) \times M_{halo}^{dyn,stack}
\label{eq:lincomb}
\end{equation}

\noindent where $a$ (equation \ref{eq:afactor}) is a linear function
of $N$ such that $a = 0$ at $N = 7$ and $a = 1$ at $N = 15$.

\begin{equation}
a = 0.125N - 0.875
\label{eq:afactor}
\end{equation}

\noindent We have chosen \mbox{$N = 15$} as our upper cutoff, as
dynamical masses for groups with \mbox{$N \geq 15$} are very
reliable. We have chosen \mbox{$N = 7$} as our lower cutoff as the
directly measured dynamical masses down to \mbox{$N = 7$} still show
roughly symmetric scatter with HAM halo masses as described in \S
\ref{sec:dynln}. For \mbox{$N < 7$}, the scatter becomes asymmetric
and we must rely on the stacked dynamical masses completely.

\begin{figure*}
\plotone{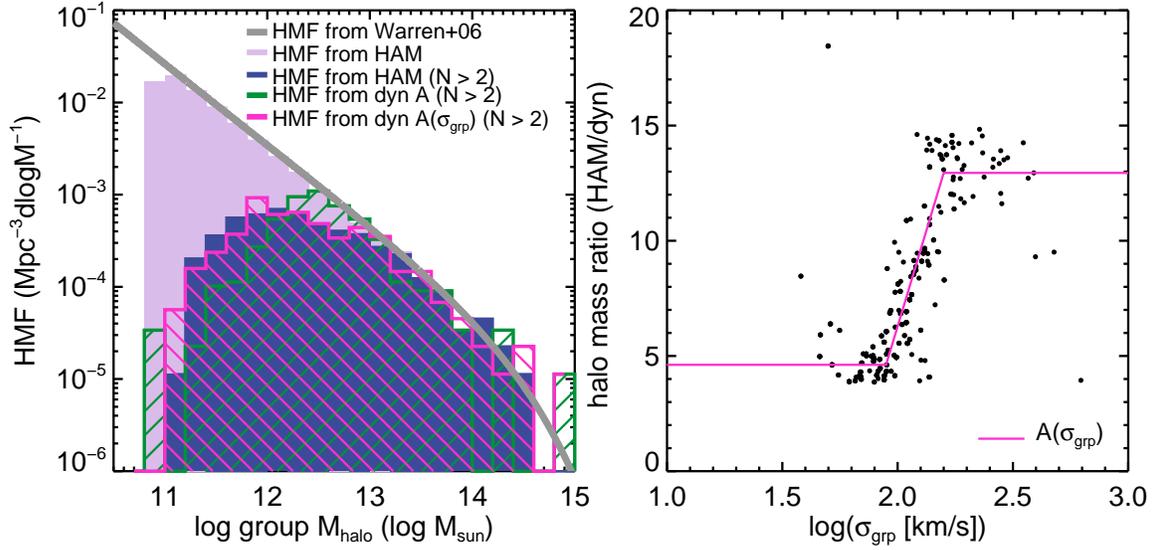}
\epsscale{1.0}
\caption{Method to determine the scale factor $A$($\sigma_{grp}$) for the
  dynamical mass estimates (using both the direct and stacking
  estimates). (a) The ECO HMF for $M_{halo}^{HAM}$ (light purple
  filled histogram) by construction matches the
  \citet{2006ApJ...646..881W} HMF (grey thick line). The ECO HMF for
  $M_{halo}^{HAM}$ for $N > 2$ groups only (dark purple filled
  histogram) is shown for comparison to the HMF of dynamical
  masses. The HMF of \mdynhat{} (scaled by a constant factor of $A$,
  green crosshatched histogram) overproduces intermediate-mass groups
  of mass $\sim$10$^{12.5}$~\msun. The HMF of \mdynhybrid{} scaled by
  $A$($\sigma_{grp}$) (pink cross-hatched histogram) is calibrated to
  reproduce the cumulative HMF for $N > 2$ groups. (b) To determine
  the scale factor A($\sigma_{grp}$), we find the HAM and dynamical
  halo masses at each group's cumulative number density and plot their
  ratio as a function of log($\sigma_{grp}$). Since we do not want
  to account for groups of $N = 1$ and $2$, we use the
  cumulative HMF from the HAM estimates and join it to the cumulative
  HMF from theory at high masses (where ECO has less data). To fit the
  data, we take the median of the halo mass ratio at high and low
  sigma (where the relationship is relatively flat) and fit a line
  between \mbox{log($\sigma_{grp}$) = 1.9--2.2} (pink). This calibration is
  applied to the dynamical mass estimates.}

\label{fg:hmf}
\end{figure*}

Before comparing with the HAM masses and addressing the low-$N$
systems further, we must calibrate our dynamical halo masses. In
Figure \ref{fg:hmf}a, we show the theoretical HMF of
\citet{2006ApJ...646..881W} as a grey line and the HMF of the ECO
$M_{halo}^{HAM}$ as a light purple histogram (matched to the
theoretical HMF by definition). We also show the $M_{halo}^{HAM}$ HMF
for $N > 2$ groups as a dark purple histogram. The \mdynhat{} HMF for
$N > 2$ groups using a constant scale factor of A is shown as the
green cross-hatched histogram, which overproduces intermediate mass
groups near $\sim$10$^{12.5}$~\msun. Since we do not necessarily think
that the characteristic group radius should stay the same as a
function of group mass, we determine a scale factor $A$($\sigma_{grp}$) that
preserves the theoretical cumulative HMF. To do this, we perform HAM
between the cumulative mass function of the raw dynamical masses and
the cumulative HMF (combining the ECO $N > 2$ M$_{halo}^{HAM}$ HMF at
low masses with the theoretical HMF at high masses). We then plot the
ratio of the abundance matched halo masses to the raw dynamical masses
as a function of group velocity dispersion. The fit to the data is
shown in pink in Figure \ref{fg:hmf}b, and we use this $A$($\sigma_{grp}$)
scale factor to create the \mdynhybrid{} HMF for $N > 2$ groups
(pink cross-hatched histogram), which better reproduces the
theoretical ($N > 2$) HMF than using the constant value of $A$. At
large halo mass, we note that the ECO dynamical mass HMF overpredicts
groups relative to the theoretical HMF. These are the few largest
clusters in ECO (including the Coma cluster) and thus their number
densities are highly subject to cosmic variance.


To incorporate HAM masses at low $N$, we construct a linear
combination of the HAM and stacked dynamical masses for groups with $3
\leq N \leq 5$, increasing the contribution from stacked dynamical
masses as a function of N. For groups with $N = 1$ and $2$, we must
rely solely on the HAM mass estimate. We note that the scatter in
group cold baryonic mass to HAM mass is $\sim$0.14 dex over the group
baryonic mass range of 10$^{10.5-11.5}$ \msun{} (after removing the
relationship with a 2nd order polynomial), while the scatter in group
cold baryonic mass to dynamical mass over the same group baryonic mass
range is $\sim$0.32 dex. The larger scatter relative to dynamical mass
is partially due to measurement uncertainties, although the smaller
scatter for HAM masses is built in due to the tightness between group
$L_r$ and cold baryonic mass (K13).  To assess the contribution from
measurement uncertainty to the dynamical mass scatter, we determine
the error on $\sigma_{grp}$ for groups with $N > 15$. We also
determine the uncertainty due to projection effects on
$\sigma_{grp}$ and $R_{proj}$ by examining the scatter in
$\sigma_{grp}$ and $R_{proj}$ at fixed $\sigma_{grp}$ and
$R_{vir}$ for known groups in a mock catalog (the same mock catalog as
used in the Appendix). By propagating these uncertainties through the
measurement of dynamical masses, we find that the typical measurement
error on the dynamical mass is $\sim$0.22 dex. In this analysis of
uncertainty, we have excluded group finding errors, which also affect
the HAM masses. Taking the quadrature difference between the measured
scatter (0.32 dex) and the measurement uncertainty (0.22 dex), we find
that the intrinsic scatter in dynamical mass is likely closer to 0.23
dex. Thus to create a smooth transition from $N = 1$ and $2$ groups to
$N \geq 3$, we match the scatter in HAM masses of $N = 1$ and $2$ groups
to 0.23 by adding 0.18 dex scatter (the quadrature sum of 0.18 and
0.14 is 0.23).




\begin{figure*}
\plotone{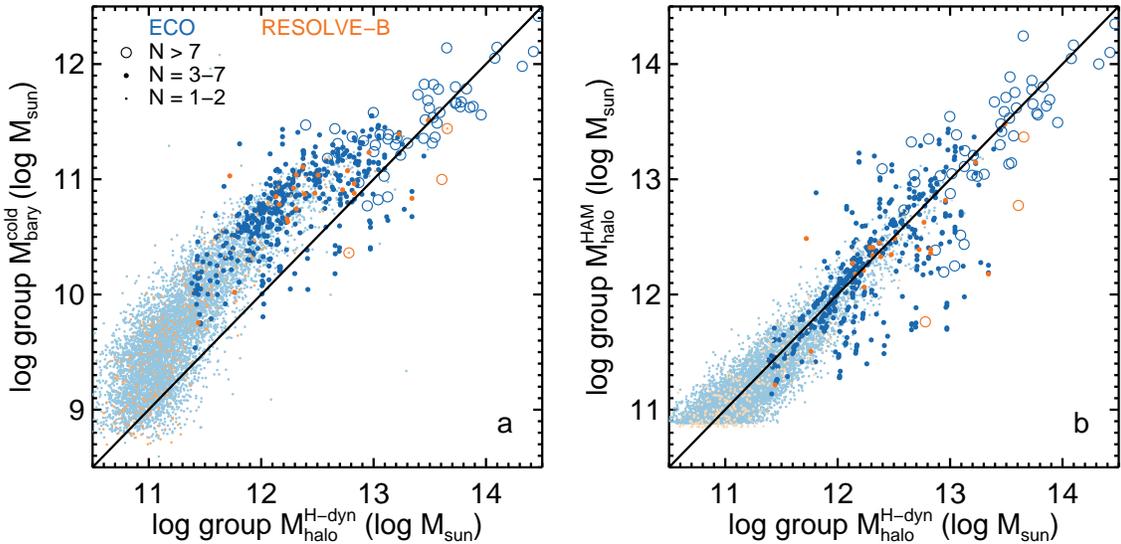}
\epsscale{1.0}
\caption{(a) Group $M_{bary}^{cold}$ (including satellite completeness
  corrections) and (b) $M_{halo}^{HAM}$ vs.\ \mdynhybrid{} using
  direct and stacking measurements at large $N$ and stacking
  measurements and HAM at low-$N$. ECO is shown in blue and RESOLVE is
  shown in orange. The black line shows a one-to-one relationship
  shifted down by 2 dex in panel (a) and a one-to-one relationship in
  panel (b). Groups with \mbox{$N > 7$} are shown as open circles,
  while groups with \mbox{$3 \leq N \leq 7$} are shown as medium sized
  dots and groups with \mbox{$N = 1$ and $2$} are shown as small
  dots. Groups are assigned mass according to group $N$ as described
  in \S \ref{sec:finaldyn}.}


\label{fg:mbvmg2}
\end{figure*}

In Figure \ref{fg:mbvmg2} we show group $M_{bary}^{cold}$ vs.\ group
\mdynhybrid{} and group $M_{halo}^{HAM}$ vs.\ group \mdynhybrid{} for
all groups. \mdynhybrid{} combines the direct and stacked dynamical
mass estimates and HAM mass estimates with scatter into one group mass
variable. We note that the curvature between halo mass and
group-integrated cold baryonic mass seen in Figure \ref{fg:mbvmg}
using the HAM group halo masses is also apparent in Figure
\ref{fg:mbvmg2}a when using the dynamical group halo masses, albeit
with larger scatter.

Examining the scatter in Figure \ref{fg:mbvmg2}b in greater detail, we
note that at fixed \mdynhybrid{} the scatter in $M_{halo}^{HAM}$
abruptly decreases below 10$^{12}$ \msun. At fixed $M_{halo}^{HAM}$,
we find that the scatter in \mdynhybrid{} below 10$^{12}$ \msun{}
is asymmetric, with greater scatter towards higher dynamical mass than
lower dynamical mass. These scatter trends highlight the limitations
of our data set as we go to lower group masses where groups have fewer
galaxies with which to accurately measure dynamical masses. They also
suggest, however, that there is greater scatter between cold baryonic
content within low-mass groups than is evident from HAM group mass
estimates.

\subsection{Group Completeness Corrections for ECO}
\label{sec:ecocc}

In \S \ref{sec:datasets} we discussed galaxy completeness corrections
computed for ECO to account for galaxies that are bright enough to be
included in our survey, but were missed due to either fiber collisions
or photometry issues. These completeness corrections are produced to
account for galaxies above our survey absolute magnitude limit. We
extend these galaxy completeness corrections to group completeness
corrections with the following simple algorithm.

The largest completeness corrections are for low-luminosity galaxies,
which are generally either satellites of larger groups or low-mass
galaxies in $N = 1$ groups. Therefore we consider two types of
completeness corrections for groups: corrections for satellites, which
affect the mass of the group, and corrections for centrals, which
affect the number density of groups.

For satellite completeness corrections, we use each satellite's galaxy
completeness correction to compute the weighted sum of stellar and
cold baryonic mass in producing group $M_{star}$ and
$M_{bary}^{cold}$. The central galaxy is automatically given a
completeness correction of 1.0 (no correction). Typically,
group-integrated stellar and cold baryonic masses are increased by
$\leq$5\%--10\%. To include these satellite completeness corrections in
the group stellar and baryonic mass distributions computed in \S
\ref{sec:grpint}, we scale the mass distribution of each satellite by
its completeness correction before performing the pseudoconvolution.


For central completeness corrections, we use the galaxy completeness
correction of each group central to weight the group mass functions
and density fields presented in \S \ref{sec:groupmfsandbfrac}. In this
case, we expect to miss $N = 1$ groups, but not large groups. Indeed,
we find that the central completeness corrections increase the number
of $N = 1$ groups by $\sim$18\%, while they increase the number of
\mbox{$N > 7$} groups by $\sim$3\%.


These group completeness corrections have not been explicitly
accounted for in the group finding, HAM, or dynamical mass
measurements. Truly accounting for the missing galaxies, however,
would affect all three. We note that increasing the number densities
of low-mass groups will systematically shift the HAM masses to lower
masses. We also note that the satellite completeness corrections
increase the group-integrated masses by $\leq$5\%--10\%, which will also
affect HAM mass estimates. For dynamical mass estimates, we can only
rely on the galaxy data available to estimate velocity dispersions.

\subsection{The Semi-analytic Model Mock Catalog}
\label{sec:sam}

To compare our results with models of galaxy evolution, we create a
mock catalog based on the SAM described in
\citet{2014MNRAS.439..264G}, which builds on the
\citet{2012MNRAS.426.2142L} model. This SAM is a variant of the
\textsc{galform} model \citep{2000MNRAS.319..168C} and is particularly
relevant to this work because it calculates separately the cold atomic
and molecular gas components, which enables both implementation of
more realistic star formation prescriptions using only the molecular
gas component \citep{2011MNRAS.418.1649L,2011MNRAS.416.1566L} and
direct comparison with our cold atomic gas data.

The \textsc{galform} model starts with the dark matter only Millennium
simulation halo merger trees \citep{2005Natur.435..629S}. This
particular SAM uses the Millennium run with the WMAP 7 cosmology
($\Omega_m = 0.27$, $\Omega_\Lambda = 0.728$, $H_0$ = 70.4~\kms;
\citealp{2011ApJS..192...18K}).  The formation and evolution of
galaxies are built on top of the dark matter only foundation by adding
gas to halos and following prescriptions for gas heating (in the form
of shocks and feedback from stars and active galactic nuclei, AGN),
gas cooling, star formation, metal enrichment, and black hole
formation. We refer the reader to \citet{2014MNRAS.439..264G} for an
in depth description of these processes. We note, however, that the
SAM used in this work differs from that of \citet{2014MNRAS.439..264G}
in its treatment of a galaxy's hot gas once it becomes a satellite
within a larger halo. In \citet{2014MNRAS.439..264G}, the satellite's
hot gas is immediately stripped upon entering the halo. In the SAM
used in this work, a ram pressure stripping algorithm described in
\citet{2014MNRAS.443.1002L} gradually removes the satellite's hot
gas. The gradual stripping of gas results in higher cold gas fractions
in early type galaxies due to the continuing accretion of gas from the
satellite's cooling hot halo, in better agreement with observations
\citep{2014MNRAS.443.1002L}. While this change in hot gas stripping of
satellites clearly affects galaxy cold baryon content, it is less
certain how it should affect the group $M_{bary}^{cold}$, as the
stripped satellite hot gas could still cool onto the central
galaxy. The cooling of gas depends on the cooling and dynamical
timescales of the (sub)halo and thus gas that may have cooled onto the
satellite may not cool onto the central galaxy. It should be noted
that the ram pressure stripping here removes the hot gas and not the
cold gas as originally described in \citet{1972ApJ...176....1G}. While
ram pressure stripping is an effective means of removing galaxy cold
gas, it primarily affects satellites of the largest clusters, of which
there are relatively few in RESOLVE and ECO.

For direct comparison to the RESOLVE-B and ECO data sets we have
produced a mock catalog from the $z = 0$ output of the SAM, by
converting the positions and velocities of its galaxies to RA, Dec,
and redshift. To compute RA and Dec, we convert from Cartesian to
spherical coordinates with the origin placed at the center of the
box. To compute redshift, we measure the distance to each galaxy from
the origin and obtain the cosmological redshift using a Hubble
constant of 70~\kms~Mpc$^{-1}$. The redshift velocity is added to the
velocity of the galaxy within the simulation, which corresponds to its
peculiar velocity.

The SAM mock catalog extends to cz = 15,000~\kms, although we cut down
the volume to a spherical shell extending between cz=2530--7470~\kms{}
and select galaxies brighter than \mbox{$M_{r}$ = $-17.33$} (similar
to ECO). To ensure that the SAM absolute magnitudes are roughly
consistent with the reprocessed magnitudes for ECO, we examine the
$r$-band luminosity function for ECO (completeness corrected) and for
the entire $z = 0$ SAM box (for greater statistics than the smaller
mock catalog). The SAM luminosity function has a shape similar to that
of ECO, albeit offset towards fainter magnitudes. Based on this
comparison, we have shifted the SAM magnitudes brighter by
$\sim$0.2~mag to be consistent with the ECO luminosity function near
\mbox{$M_{r,tot}$ = $-23.0$}, where the ECO luminosity function
reaches 10 galaxies per Mpc$^{3}$. The final mock volume is
$\sim$1649480 Mpc$^{3}$ and has an overall number density of galaxies
brighter than $-17.33$ of $\sim$0.0233~Mpc$^{-3}$ (similar to the
completeness corrected ECO number density of $\sim$0.0247~Mpc$^{-3}$).

We perform FOF group finding and HAM for the SAM using the same codes
as described in \S \ref{sec:envandhmass} so that we can examine both
the ``true'' groups and the ``FOF'' groups that would be identified by
an observer with group finding errors. We perform the same algorithm
to break up false pairs as was used for the ECO and RESOLVE-B data
sets. To reduce errors in group finding due to galaxies with large
peculiar velocities, we further cut down the mock to galaxies with
\textit{group redshifts} within a spherical shell 3000--7000~\kms{}
(similar to ECO). Within this smaller volume, there are 65,784 true
groups (57,109 are \mbox{$N = 1$}) and there are 69,161 FOF groups
(58,587 are \mbox{$N = 1$}).

For both the true and FOF groups we have measured the group $M_{star}$
and $M_{bary}^{cold}$. We note that the IMF used in the model is that
of \citet{1983ApJ...272...54K}, different from the
\citet{2003PASP..115..763C} IMF used to compute stellar masses for
RESOLVE-B and ECO. The two IMFs yield similar mass-to-light ratios, so
we do not expect this difference to cause significant systematics in
comparing the model and the data. For cold baryonic mass, we sum the
stellar and cold atomic gas mass in each galaxy, for which cold atomic
gas mass is defined as 1.4$M_{HI}$ to account for helium. While the
SAM records the exact amount of HI mass in each galaxy, our
observations frequently allow us to constrain the gas mass as an upper
limit $\sim$5\%--10\% of the galaxy's stellar mass. To reflect this
observational effect on the SAM, we replace the gas mass with
0.05$M_{star}$ if its value is less than 5\% of the stellar mass.


To ensure that the SAM true and FOF group halo masses are consistent
with the ECO data, we compare the HMFs. We find no significant offset
and thus we apply no correction to the group halo masses.

\section{Group Mass Functions and Baryon Fractions}
\label{sec:groupmfsandbfrac}

We now examine the group-integrated stellar and cold baryonic mass
functions (or group SMF and CBMF) and group-integrated stellar and
cold baryon fractions for the RESOLVE-B and ECO data sets as well as
for the SAM mock catalog.

\subsection{Group Mass Functions}
\label{sec:groupmfs}

To measure the group SMF and CBMF for RESOLVE-B and ECO, we adapt the
cross-bin sampling technique described in E16. For the galaxy MFs in
E16, this method first combines all individual mass likelihood
distributions into one combined survey mass likelihood distribution by
summing the likelihoods in each bin. Then the overall stellar or
baryonic mass functions are constructed by sampling from the combined
survey mass likelihood distribution 1000 times in a Monte Carlo
fashion. From these 1000 samples, we determine the median and the
uncertainty bands (16th-84th percentiles of the mass functions). In
this work, rather than constructing the mass likelihood distribution
of all the galaxies in our data set, we construct the mass likelihood
distribution of all the groups in our data set, using the individual
group mass likelihood distributions computed in \S
\ref{sec:grpint}. As in E16, the likelihoods for ECO are weighted by
the central completeness correction factor (described in \S
\ref{sec:ecocc}).

\begin{figure*}
\plottwo{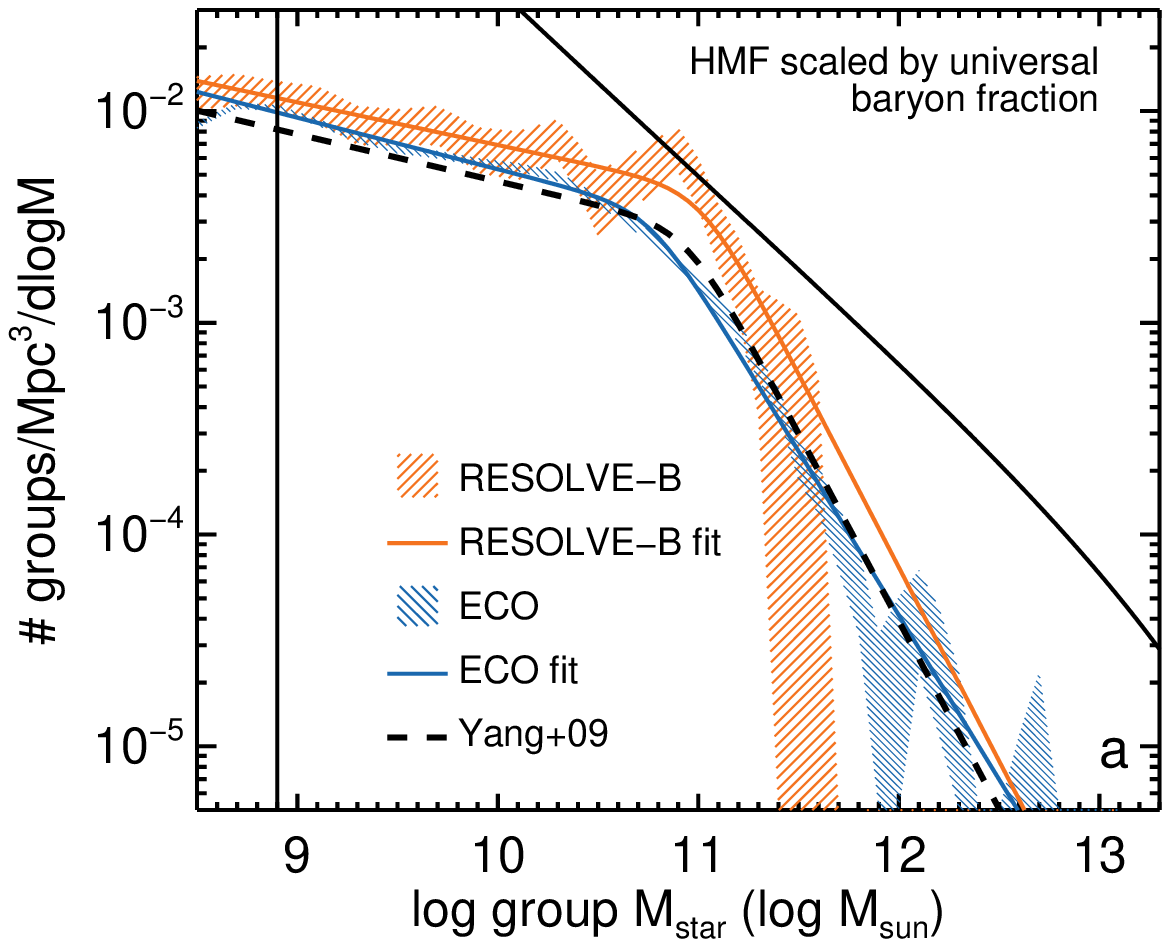}{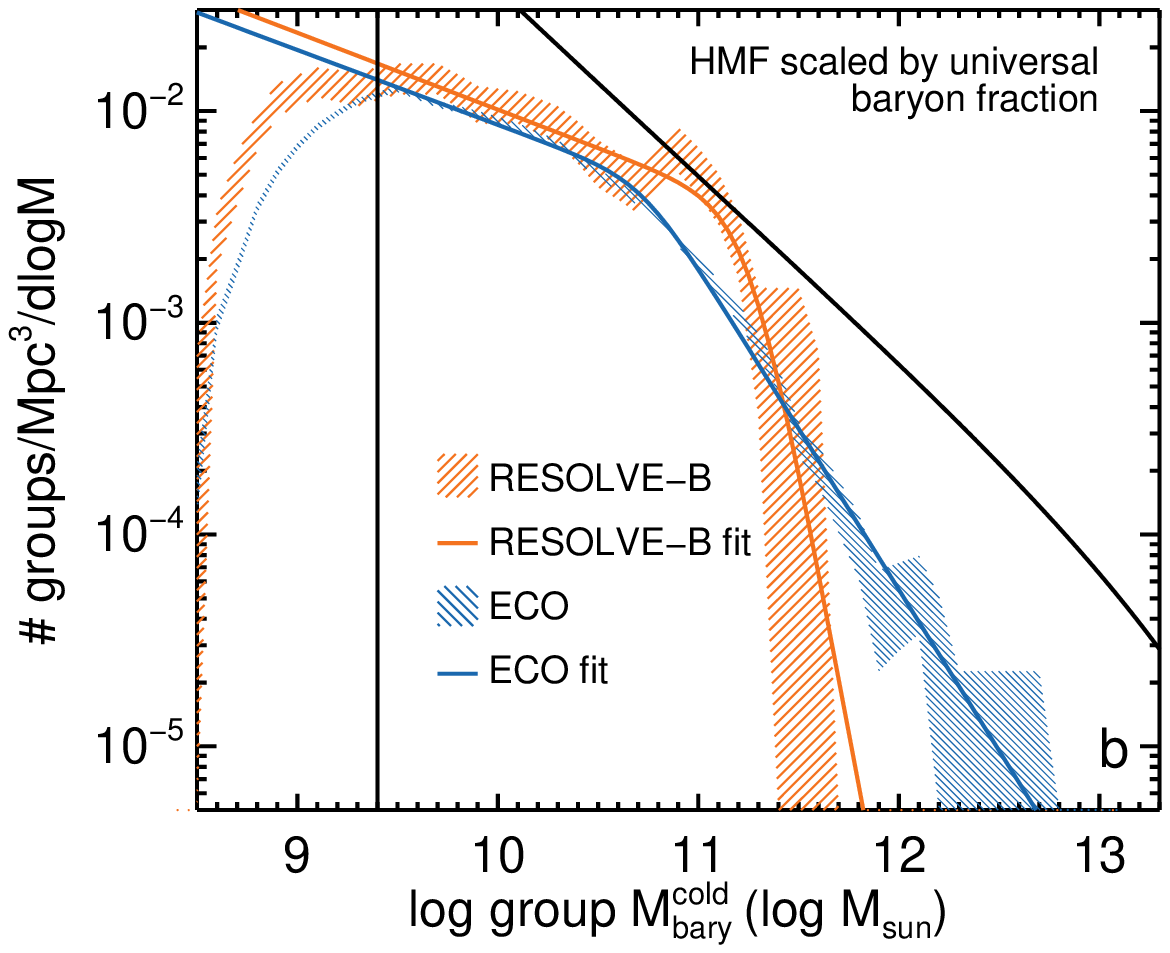}
\epsscale{1.0}
\caption{Group (a) SMF and (b) CBMF for ECO (blue) and RESOLVE-B
  (orange) determined using the cross-bin sampling method of E16. The
  cross-hatched regions show the 16th-84th percentile uncertainty
  bands and the black vertical lines designate the group $M_{star}$
  and $M_{bary}^{cold}$ completeness limits. The theoretical dark
  matter HMF of \citet{2006ApJ...646..881W} scaled by a universal
  baryon fraction of 0.15 \citep{2014A&A...571A..16P} is shown in
  black. In both panels the group MFs fall short of the theoretical
  HMF at both high and low masses. The black dashed line in panel (a)
  shows the double power law fit to the group SMF given in
  \citet{2009ApJ...695..900Y}. The dark blue and orange lines in both
  panels show our fits using the same model for the ECO and RESOLVE
  group MFs. The group CBMF shows similar behavior to the
  group SMF at high masses and a steeper low-mass slope, as also seen
  in the galaxy mass function due to the abundance of gas-dominated
  galaxies at low halo masses (E16).}

\label{fg:smfbmf}
\end{figure*}

The group SMFs for RESOLVE-B and ECO are shown in Figure
\ref{fg:smfbmf}a. The fact that RESOLVE-B is elevated over ECO is due
to cosmic variance. In E16, we found that RESOLVE-B is overdense at
intermediate and low galaxy masses, although underdense at large
masses due to a lack of cluster sized halos, compared to ECO. Figure
\ref{fg:smfbmf}a also compares the group SMF with the HMF from
\citet{2006ApJ...646..881W}, which does not include the contribution
from subhalos and is scaled by a universal baryon fraction of 0.15
\citep{2014A&A...571A..16P} for direct comparison with the data. We
also use this HMF to assign HAM halo masses to the data and SAM mock
catalog.


At high masses, the group SMF drops off more steeply than the
universal baryon fraction scaled HMF. Around M$_{star}$ = 10$^{10.9}$
\msun{} the group SMF reaches a maximum compared to the HMF, which we
determine by finding the group $M_{star}$ at the maximum of the ratio
between the group SMF and the universal baryon fraction scaled
HMF. Below 10$^{10.9}$~\msun, the group SMF exhibits a shallow rise,
even shallower than the galaxy SMF (not shown, see E16). This slow
rise reflects the fact that low-mass satellites from the galaxy SMF
are removed from the low-mass end and placed into high-mass groups in
the group SMF. This result clearly illustrates the large gap between
the group SMF and dark matter HMF at low group-integrated stellar
masses.

To compare with previous work, we overplot the double power-law fit to
the group SMF from \citet{2009ApJ...695..900Y} in Figure
\ref{fg:smfbmf}a as a black dashed line. They used the
\citet{2007ApJ...671..153Y} group catalog constructed from the
NYU-VAGC SDSS galaxy catalog \citep{2005AJ....129.2562B}. Their group
finding algorithm used FOF to define potential groups and then an
iterative process to assign galaxies to each potential group based on
its mass and size. To fit their group SMFs, they used the equation:

\begin{equation}
\phi(m)=\phi_{*}\frac{(\frac{m}{M_{*}})^{\alpha}}{(x_0+(\frac{m}{M_{*}})^4)^{\beta}}
\label{eq:yang}
\end{equation}

\noindent and found parameters: $\phi_{*} = 0.00731$, $\log{M_{*}} =
10.67$, $x_0 = 0.7243$, $\alpha = -0.2229$, and $\beta = 0.3874$.

We have performed fits using a modified version of this double power
law model, fixing $x_0 = 1$ since this parameter is not well
constrained by our data. We find that fixing $x_0$ affects the values
of $\phi_{*}$ and $\log{M_{*}}$ but not $\alpha$ and $\beta$. Our own
fits using this modified double power law model are shown as solid
orange and blue lines for RESOLVE-B and ECO respectively and the model
parameters are given in Table \ref{tb:mffits}. We find a similar
low-mass slope for both RESOLVE-B and ECO of $\alpha \sim -0.2$ to
that found in \citet{2009ApJ...695..900Y}. At higher group masses, we
observe slight differences in the group SMF knee and high-mass falloff
between ECO and the fit from \citet{2009ApJ...695..900Y}, though these
are likely within systematic uncertainties between studies such as the
stellar mass estimation and cosmic variance. For RESOLVE-B, the double
power law model does not fit the data well at high masses, as can be
seen by the fit as well as the large errors on the $\beta$ parameter,
which describes the power law slope of the high mass end. The poor fit
at high masses is likely due to the fact that RESOLVE-B is a small
volume (i.e., highly subject to cosmic variance) and has no large
clusters of mass $>10^{13.5}$~\msun{} (E16), resulting in a very steep
dropoff at high group masses. It was further shown in E16 that it is
possible to reconstruct the RESOLVE-B \textit{galaxy} mass function by
scaling a set of basis conditional mass functions (galaxy mass
functions broken down into different halo mass regimes) by the number
of group halos in each halo mass regime, thus suggesting that a given
survey's mass function can be determined by its halo mass
distribution.


\begin{deluxetable*}{ccccc}
\tablecaption{Fitting Parameters for Group Integrated Mass Function}
\startdata
\tablehead{\colhead{data} & \colhead{$\phi_{*}$} & \colhead{log$M_{*}$} & \colhead{$\alpha$} & \colhead{$\beta$} \\ \colhead{} & \colhead{dlogM$^{-1}$Mpc$^{-3}$} & \colhead{log(M$_{\odot}$)} & \colhead{} & \colhead{} } 
RESOLVE-B group SMF & 0.0043 $\pm$ 0.0011 & 11.02 $\pm$ 0.28 & -0.20 $\pm$ 0.05 & 0.41 $\pm$ 0.31 \\
ECO group SMF & 0.0035 $\pm$ 0.0002 & 10.76 $\pm$ 0.04 & -0.24 $\pm$ 0.01 & 0.33 $\pm$ 0.02 \\
RESOLVE-B group CBMF & 0.0035 $\pm$ 0.0018 & 11.26 $\pm$ 0.46 & -0.36 $\pm$ 0.07 & 1.17 $\pm$ 3.25 \\
ECO group CBMF & 0.0048 $\pm$ 0.0003 & 10.73 $\pm$ 0.04 & -0.35 $\pm$ 0.02 & 0.29 $\pm$ 0.02 \\
\enddata
\label{tb:mffits}
\end{deluxetable*}

The group CBMFs for RESOLVE-B and ECO are shown in Figure
\ref{fg:smfbmf}b. At high mass, they are very similar to the group
SMFs, since high mass groups generally have little cold gas (e.g.,
\citealp{1984ARA&A..22..445H}). Indeed we find that the power-law
slopes of the high-mass ends are very similar between the group SMF
and CBMF (\mbox{$\beta$ $\sim$ 0.3} for ECO; we note that RESOLVE-B is
not well constrained at high masses). At $M_{bary}^{cold}$ =
10$^{10.9}$~\msun{} the group CBMF also reaches its relative maximum
to the group HMF (measured again by finding the group
$M_{bary}^{cold}$ at the maximum of the ratio of the group CBMF to the
scaled HMF). Below 10$^{10.9}$~\msun, the group CBMF rises more
steeply than the group SMF (\mbox{$\alpha_{SMF} \sim -0.24$}
vs.\ \mbox{$\alpha_{CBMF} \sim -0.35$}), similar to the galaxy BMF
vs.\ SMF shown in E16. However, the group CBMF is still not as steep
as the HMF.

At all masses, the group SMF and CBMF fall below the scaled HMF,
although perhaps for different reasons. In large groups, we expect
that hot gas dominates the baryon content, and we will examine its
effect on the group BMF in \S \ref{sec:discussaddhotgas}. At lower
masses, the separation from the HMF may also be indicative of
unaccounted for gas (see \S \ref{sec:undgas}) or due to other
processes related to galaxy and group formation.

\begin{figure*}
\plottwo{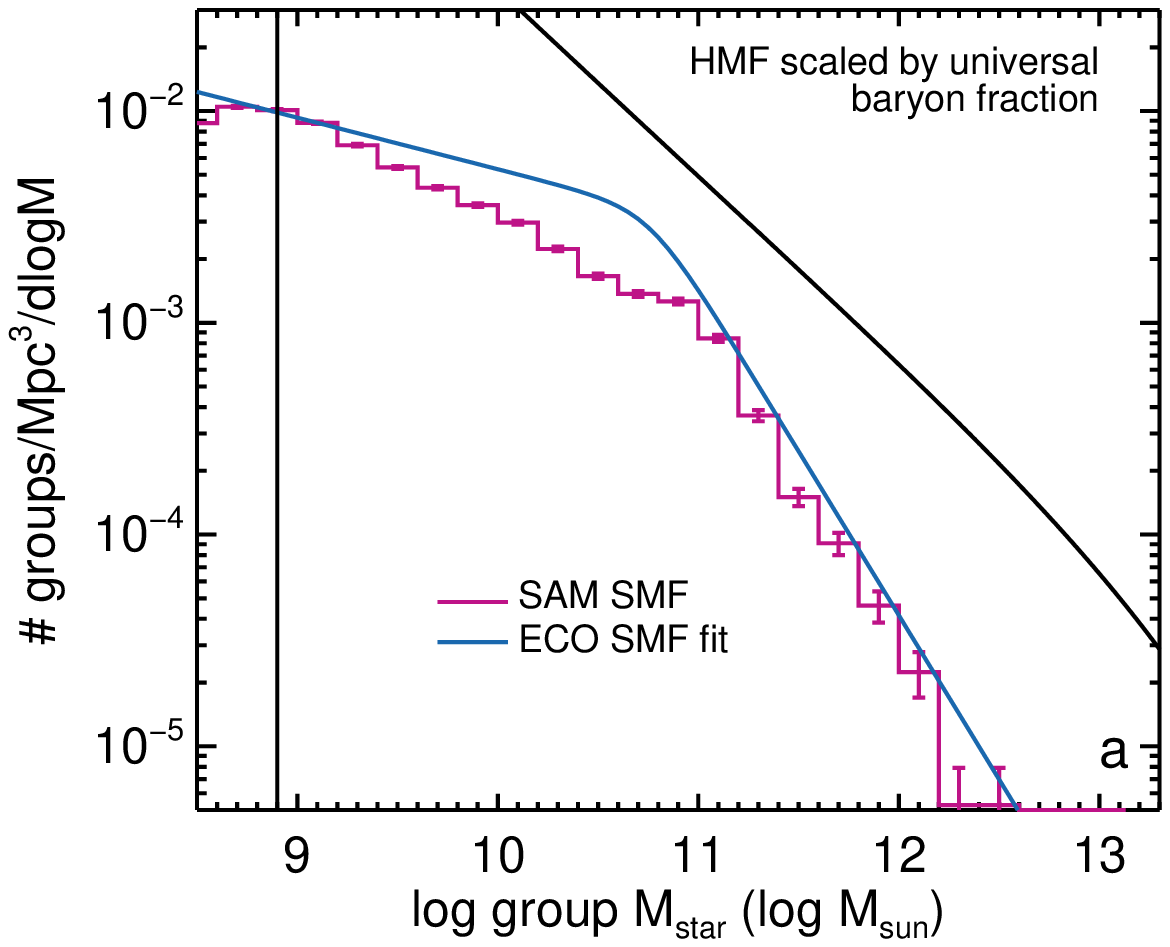}{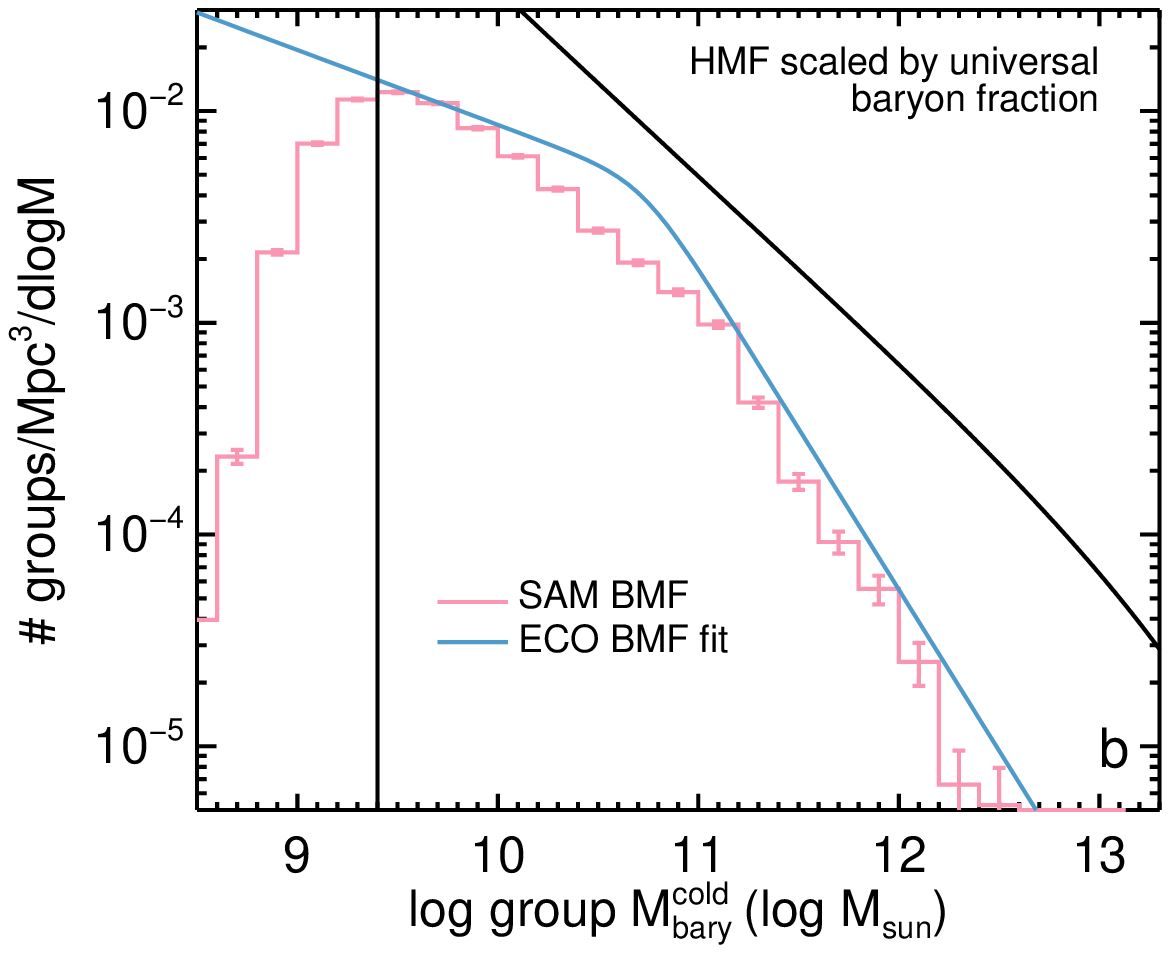}
\epsscale{1.0}
\caption{Group (a) SMF and (b) CBMF for the FOF groups in the GALFORM
  semi-analytic model. (Results are not shown for the mock catalog
  true groups but are similar.) Since the simulation's stellar and
  baryonic mass are exact quantities, we plot the mass functions with
  Poisson error bars only. The theoretical dark matter HMF scaled by
  the universal baryon fraction is shown in black and the ECO group
  SMF and CBMF are also shown plotted in their respective panels. The
  SAM and ECO group MFs exhibit similar overall behavior, with a break
  in power-law slope near $\sim$10$^{11}$~\msun. The SAM, however,
  shows a dip in numbers relative to ECO at this mass scale and a more
  steeply rising slope at the lowest masses. The solid vertical lines
  show the stellar and cold baryonic completion limits of the ECO
  survey.}

\label{fg:samsmfbmf}
\end{figure*}

In Figure \ref{fg:samsmfbmf}, we show the group SMF (magenta) and
CBMF (light pink) for the FOF groups in the SAM mock catalog. We note
that the results do not change significantly if we use the true groups
rather than the FOF groups. Since the galaxy masses are known to
precision (albeit with systematic errors due to the assumptions in the
model such as the IMF, star formation prescription, and gas
definition), we bin the values and assume Poisson error bars for each
bin.

At high mass ($>$10$^{11}$~\msun), the group SMF and CBMF for the SAM
are similar to the ECO group SMF and CBMF (overplotted in dark and
light blue respectively). Below $\sim$10$^{11}$~\msun, however, the
SAM group SMF and CBMF deviate significantly from the data. We find
that the SAM contains many fewer groups with group
\mbox{$M_{bary}^{cold}$ $\lesssim$ 10$^{11}$~\msun{}}. The knee of the
SAM group SMF and CBMF is located at higher masses than that observed
in ECO. We explore the location of the knee of the group SMF and CBMF
and its relation to the feedback implementation in the model in more
detail in \S \ref{sec:nascentgroups}. The SAM group SMF and CBMF then
rise more steeply at the lowest group halo masses.




\subsection{Group Stellar \& Cold Baryon Fractions}
\label{sec:grpbfracs}

\begin{figure*}
\plottwo{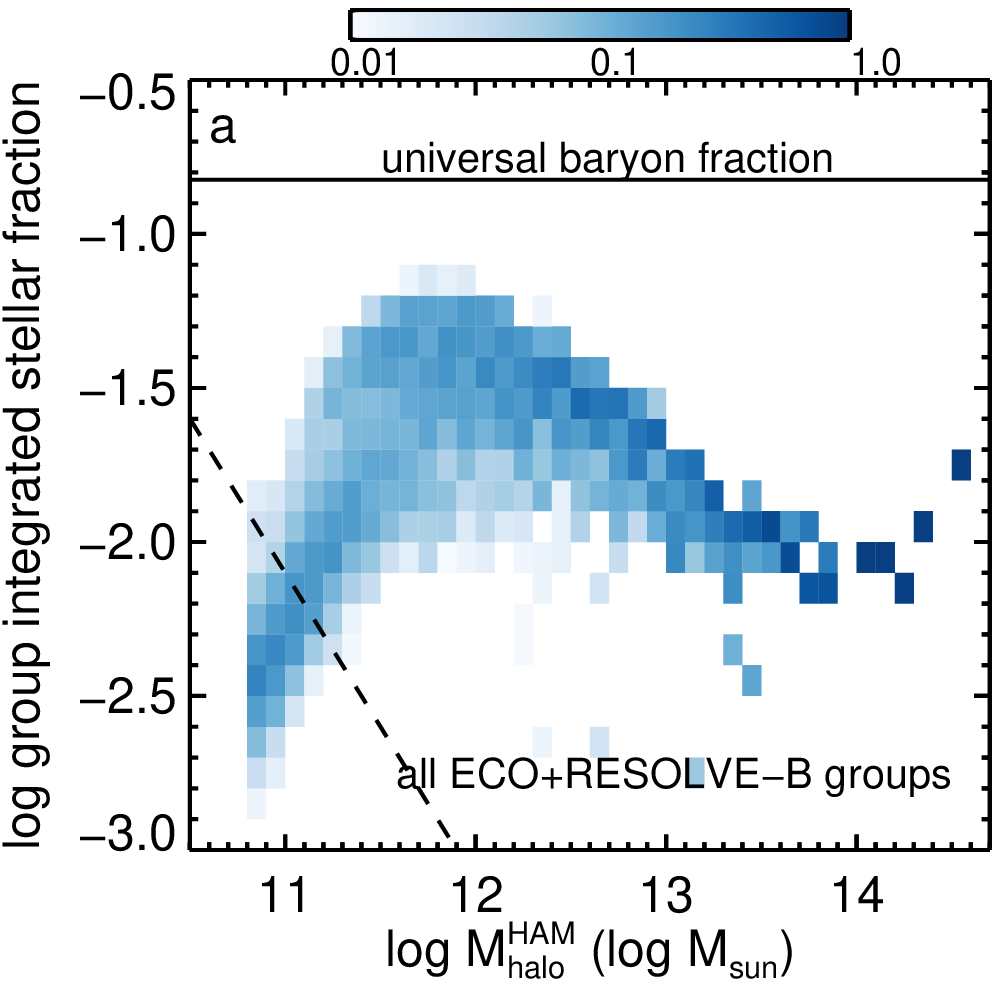}{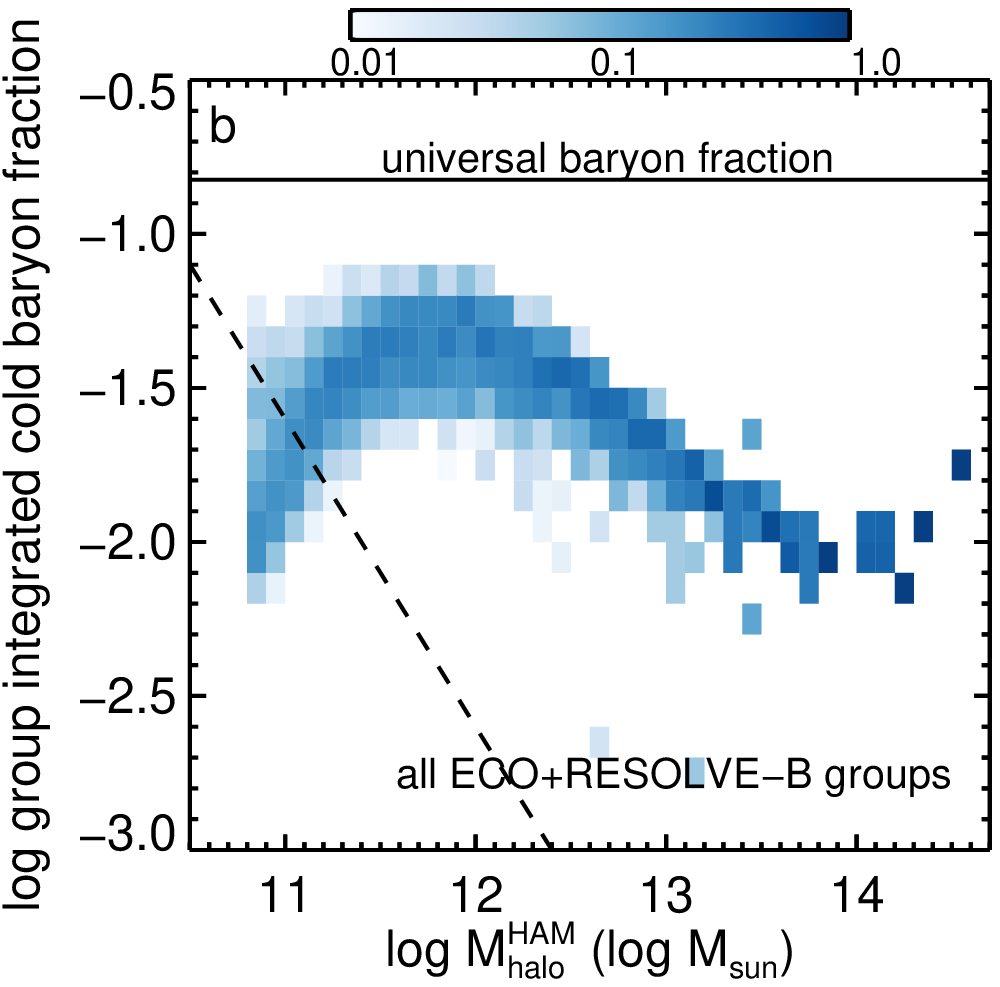}
\epsscale{1.0}
\caption{Conditional density plot of group-integrated (a) stellar and
  (b) cold baryon fraction as a function of $M_{halo}^{HAM}$ based on
  group $L_r$ for all ECO and RESOLVE-B groups. The number
  densities are weighted by the central completeness corrections
  described in \S \ref{sec:ecocc}.  The universal baryon fraction
  (0.15) is shown as a line at the top and the group $M_{star}$ and
  M$_{bary}^{cold}$ completeness limits are shown by the dashed black
  lines. The group-integrated stellar mass fraction falls sharply
  above and below group halo mass $\sim$10$^{11.8}$~\msun. The
  group-integrated cold baryon fraction reaches a maximum over the
  nascent group regime (10$^{11.4-12}$~\msun) with little scatter.  At
  high group halo mass, the stellar and cold baryon fractions are
  similar. }
\label{fg:bfrac}
\end{figure*}

\begin{figure*}
\plottwo{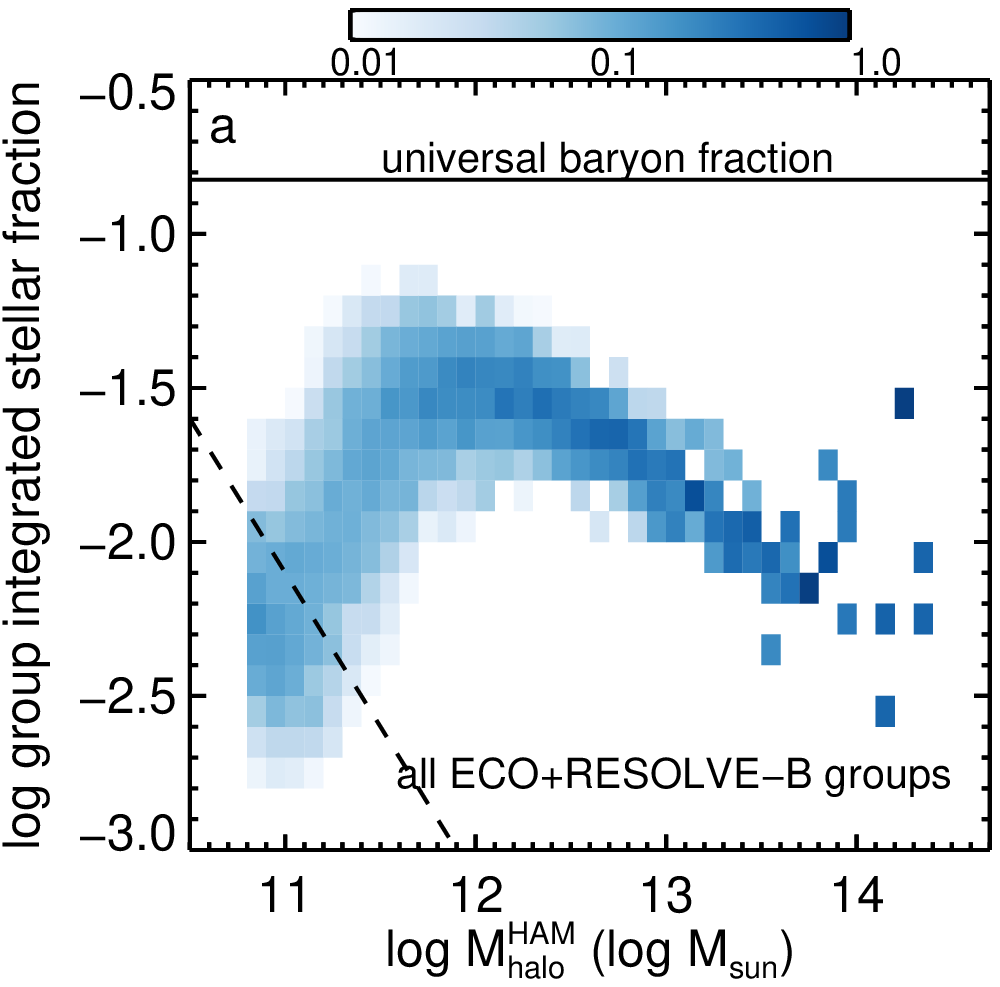}{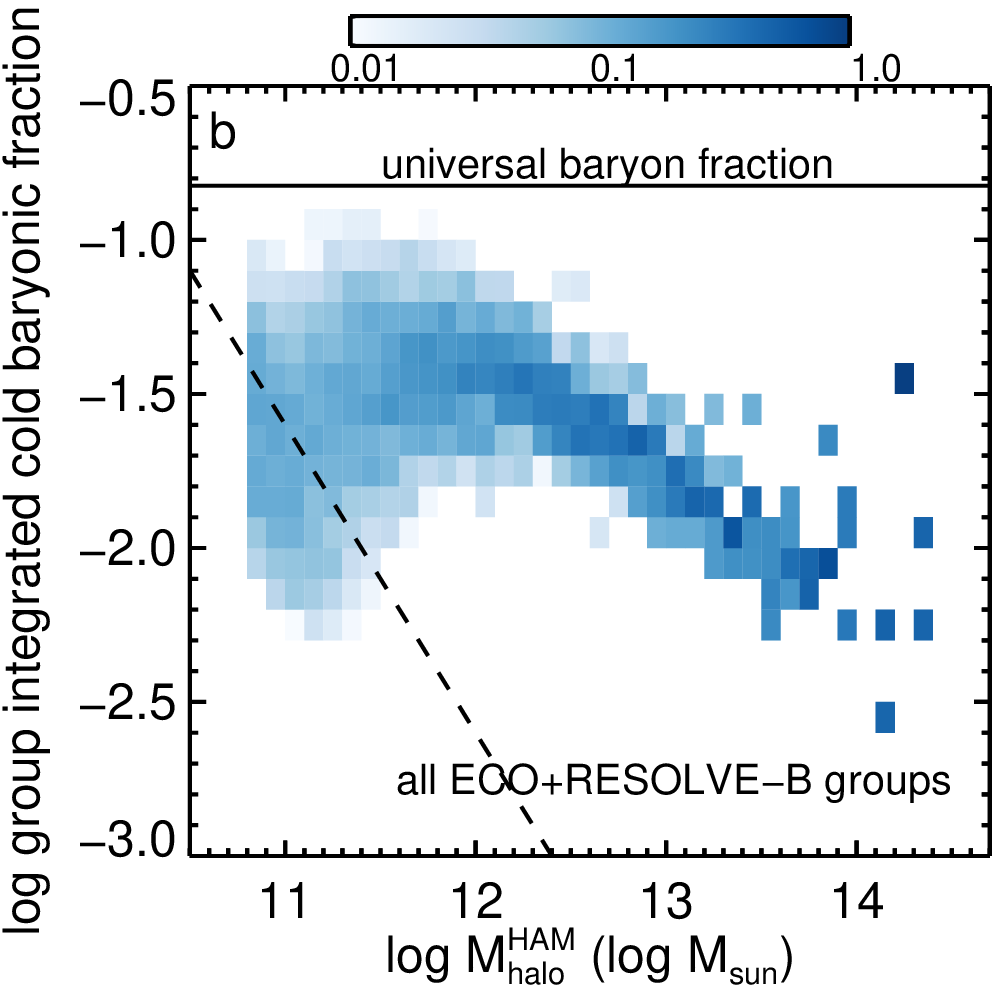}
\epsscale{1.0}
\caption{Same as Figure \ref{fg:bfrac}, except using $M_{halo}^{HAM}$
  based on group $M_{star}$ abundance matching with scatter of 0.14
  dex added (to match the scatter based on group luminosity). The
  group-integrated stellar fractions are similar to those using the
  group masses matched on $L_r$. The group-integrated cold baryon
  fractions have increased scatter at low masses compared to the $L_r$
  version. The increased scatter is likely due to the fact that $L_r$
  correlates more closely with cold baryonic mass than with stellar
  mass.}
\label{fg:bfrac2}
\end{figure*}

\begin{figure*}
\plottwo{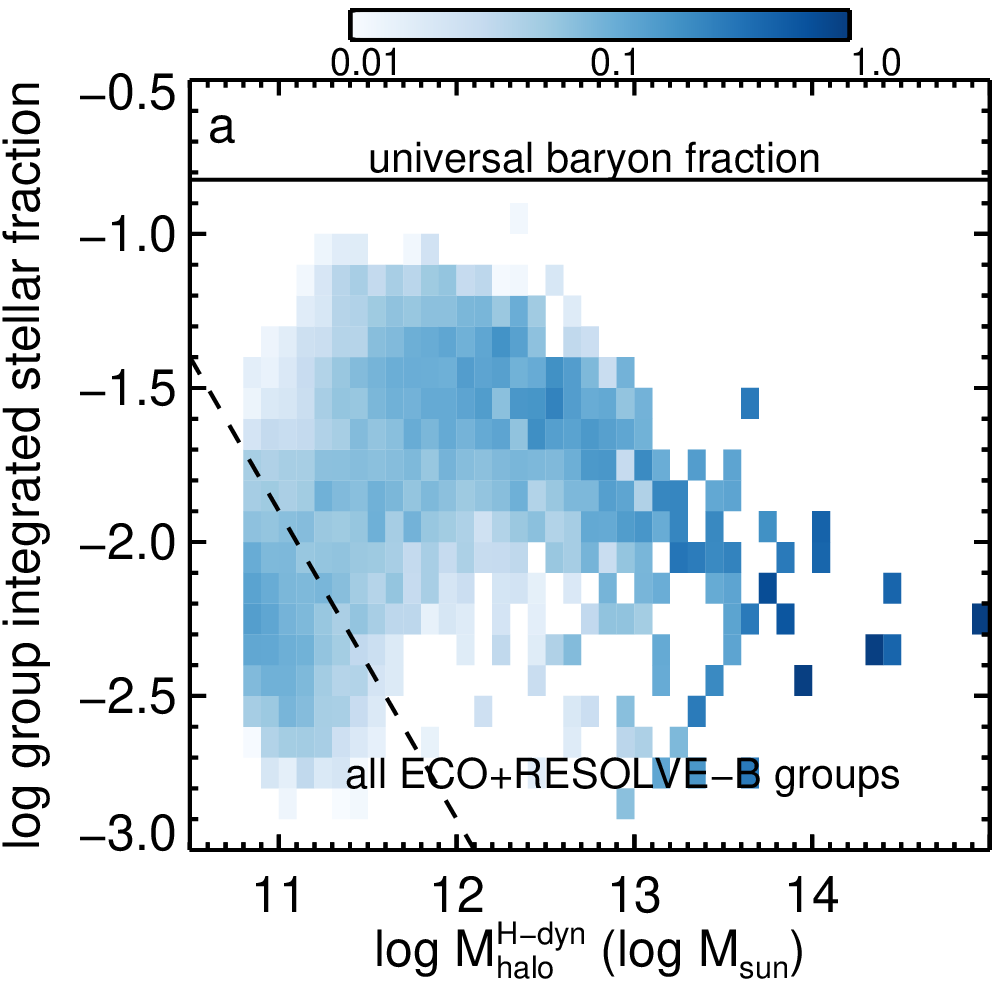}{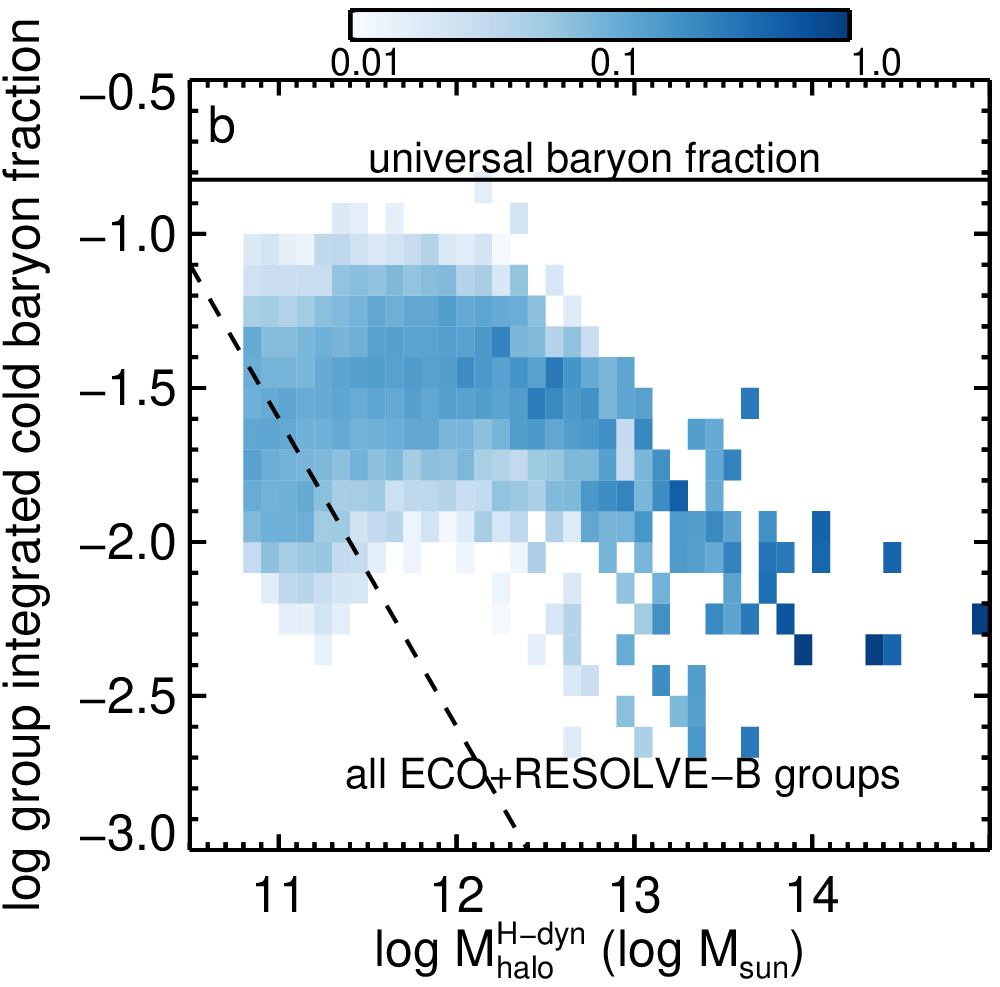}
\epsscale{1.0}
\caption{Conditional density plot of group-integrated (a) stellar and
  (b) cold baryon fraction as a function of \mdynhybrid{} determined
  by the combination of $N \geq 3$ dynamical and low-N HAM estimates
  for both ECO and RESOLVE-B groups combined. The universal baryon
  fraction is shown as a line at the top and the group cold baryonic
  mass completeness limit is shown by the dashed black line. Using
  dynamical mass results in a wider range of group stellar and cold
  baryon fractions, particularly over the nascent group halo mass
  range (and below, but here the scatter is introduced see \S
  \ref{sec:finaldyn}). The peak baryonic collapse efficiency occurs
  near $\sim$10$^{12}$~\msun. }

\label{fg:bfracdyn}
\end{figure*}

We now examine group-integrated stellar and cold baryon fractions as a
function of halo mass for ECO and RESOLVE-B. These fractions are
defined as the group $M_{star}$ or $M_{bary}^{cold}$ divided by
$M_{halo}$. We can interpret these fractions as the stellar or
baryonic collapse efficiency of groups, i.e., how many of the group
halo baryons have collapsed into the observable stars and cold gas in
galaxies. To increase our statistics we analyze the ECO and RESOLVE-B
data sets together, using the group central completeness corrections
to weight ECO groups appropriately (see \S
\ref{sec:ecocc}).\footnote{We do not find any offsets between the two
  surveys when analyzed separately.}  Figures \ref{fg:bfrac},
\ref{fg:bfrac2}, and \ref{fg:bfracdyn} compare results using HAM and
dynamical group mass estimates. The group-integrated stellar and cold
baryonic mass completeness limits are shown as dashed black lines.

Using HAM (Figures \ref{fg:bfrac} and \ref{fg:bfrac2} show results
based on matching on group $L_r$ and group $M_{star}$ respectively),
we find that the group stellar and cold baryon fractions peak around a
halo mass of $\sim$10$^{11.8}$~\msun, although the cold baryon
fraction peak is much broader than that of the stellar fraction.  At
higher and lower group halo mass, the stellar and cold baryon
fractions fall off. This behavior has been seen in previous work
examining the stellar fraction using HAM or HOD methods for assigning
halo mass (e.g., \citealp{2012ApJ...746...95L}). The falloff towards
higher group halo masses is interpreted to show that high-mass groups
are increasing their hot gas and dark matter content faster than their
collapsed cold baryonic mass content. The falloff toward low group
halo masses would then correspondingly show that low-mass galaxies
(generally single galaxies in their own halo down to our survey
limits) are growing rapidly in cold baryonic mass (from cooling halo
gas), while their halo mass is not increasing as quickly
\citep{2009ApJ...696..620C,2010ApJ...710..903M}. We note that in the
largest group mass bin the stellar and cold-baryon fraction
rises. This bin, however, consists of one group, whose corresponding
dynamical mass does not confirm such a rise as physically meaningful
(Figure \ref{fg:bfracdyn}).

An alternative, static interpretation of these plots is that low-mass
galaxies have suppressed cold baryonic content due to stellar and
supernova feedback. Studies based on simulations suggest, however,
that dwarf galaxies with gas masses of 10$^{8-9}$~\msun{} do not lose
their gas due to stellar and supernova feedback
\citep{1999ApJ...513..142M,2015MNRAS.446..299M}, and in addition
cooling of fresh and recycled gas in low-mass halos is generally found
to be efficient \citep{2011MNRAS.416..660L,2016arXiv161008523A}. In
fact, the isolated dwarfs in this regime are far from static, as they
are doubling their stellar masses on $\sim$Gyr timescales
(K13). Reionization at early times may have heated the gas in
lower-mass halos, thus delaying their formation relative to higher
mass galaxies, but at present their growth rates are high. We also
note that galaxies below the survey limit could contribute up to
$\sim$14\% of the cold baryonic mass in low-mass groups
(\mbox{$M_{halo}$ $<$ 10$^{11.4}$ \msun}, see \S \ref{sec:grpint}),
increasing the cold baryon fractions by roughly $\sim$0.05 dex.


In Figure \ref{fg:bfrac} using HAM based on group luminosity, the peak
of the stellar mass fraction is sharper than that of the cold baryonic
mass fraction, with a steeper falloff towards lower mass
groups. Adding the cold gas results in a flatter peak over the nascent
group regime that has little scatter in cold baryon fraction. In
Figure \ref{fg:bfrac2} using HAM based on group $M_{star}$, however,
we find that the group cold baryon fraction becomes much more
scattered at low masses, likely due to the fact that stellar mass HAM
does not track cold baryonic mass as well as group $L_r$ HAM does in
the low group halo mass regime.

In Figure \ref{fg:bfracdyn}, we show the group stellar and cold baryon
fractions as a function of dynamical mass as described in \S
\ref{sec:finaldyn}. At high group halo masses, the stellar and cold
baryon fractions are similar to results using HAM, decreasing with
increasing halo mass. At all masses, however, we find a much greater
diversity in stellar and cold baryon fractions with \mdynhybrid{} than
with $M_{halo}^{HAM}$ (e.g., width $\sim$1~dex vs.\ $\sim$0.5~dex near
\mbox{$M_{halo}$ = 10$^{12}$ \msun}). Thus HAM may build in a
perceived tight maximum in baryonic collapse efficiency over the
nascent group regime, whereas the dynamical masses suggest there
should be more scatter, potentially due to variations in the hot gas
fractions within groups. We note that Figure \ref{fg:bfracdyn}
includes $N = 1$ and $2$ groups, which rely purely on HAM (with
increased scatter to smoothly transition to \mbox{$N \geq 3$}
dynamical masses) for mass estimates.

\begin{figure*}
\plottwo{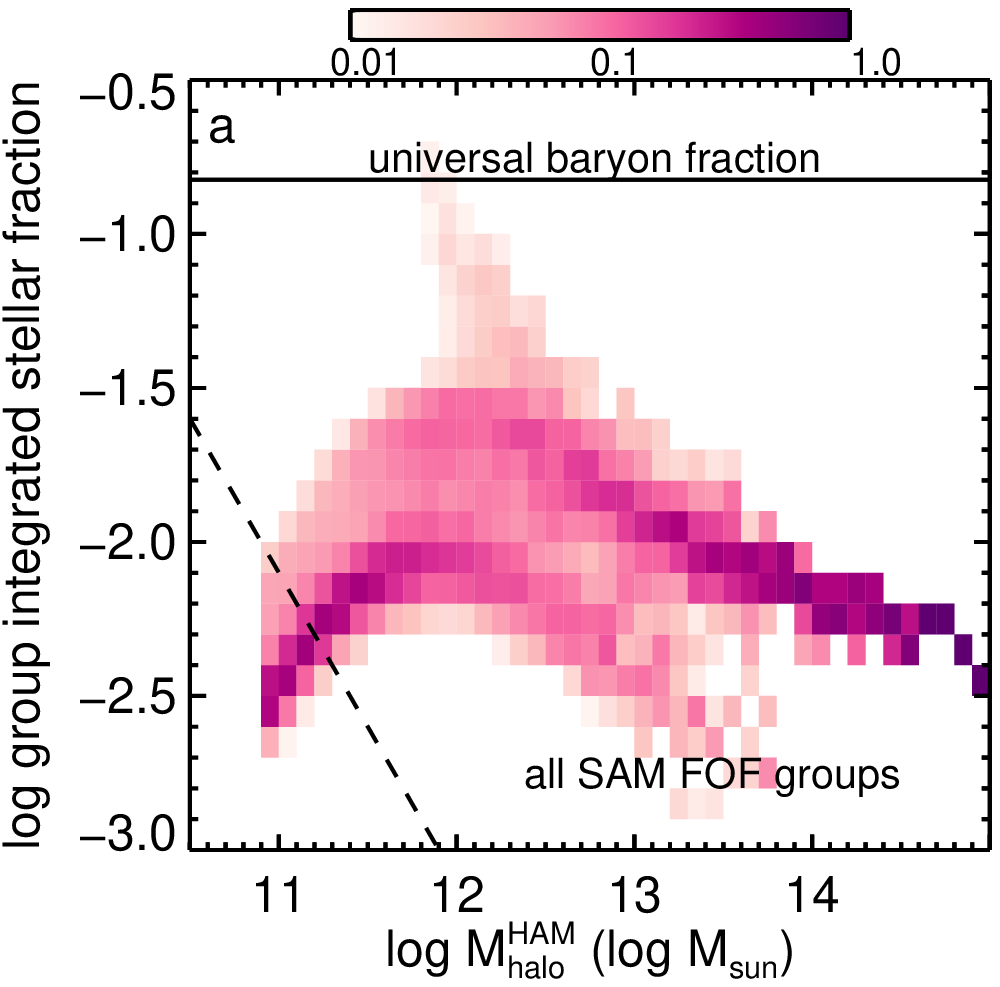}{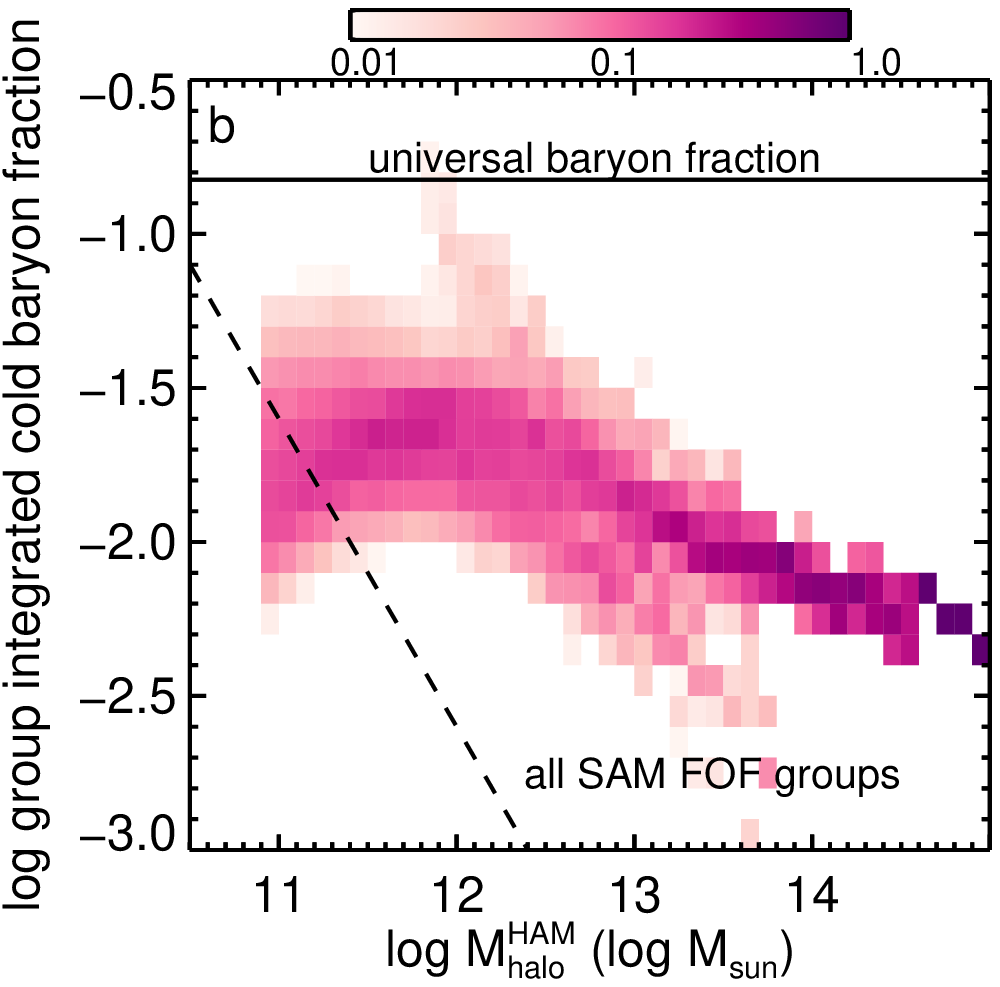}
\epsscale{1.0}
\caption{Conditional density plot of group-integrated (a) stellar and
  (b) cold baryon fraction for the SAM using the FOF groups with HAM
  based on group luminosity. We find that applying FOF and HAM to the
  SAM mock catalog results in the upside down U shape as seen in the
  ECO data. The spur of low cold baryon fraction groups for their halo
  mass is due to a population of galaxies with much lower masses than
  their luminosities would suggest. These are large, blue, star-forming
  galaxies, for which the dust correction does not sufficiently
  extinct their intrinsic magnitudes (see discussion in Gonzalez-Perez
  et al. submitted).}

\label{fg:samfracham}
\end{figure*}
\begin{figure*}
\plottwo{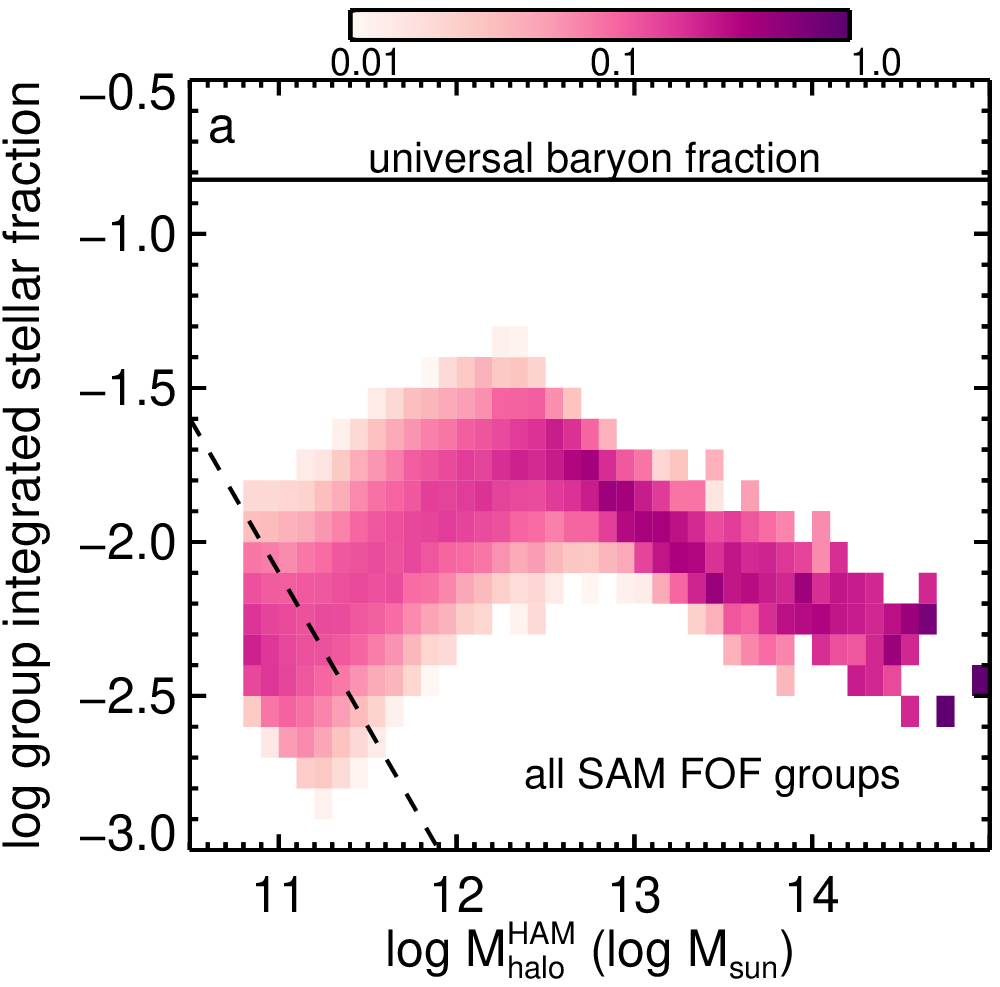}{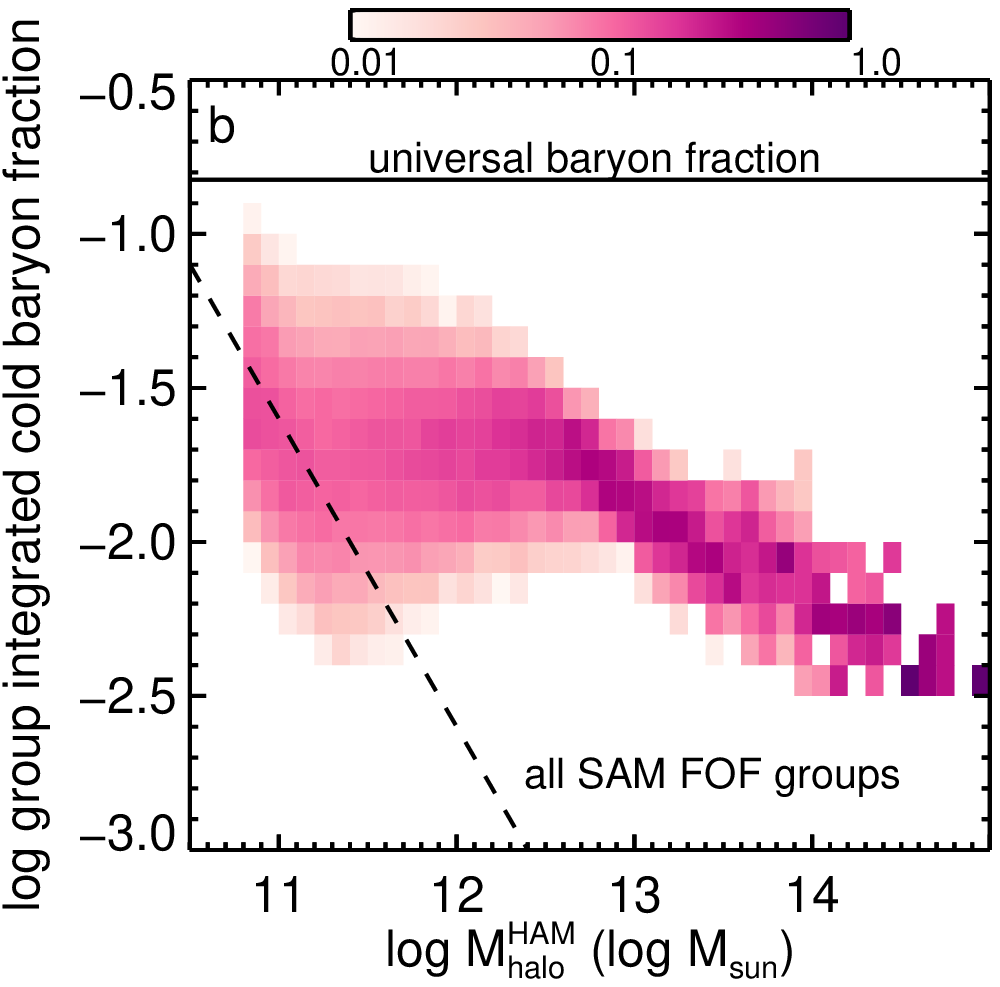}
\epsscale{1.0}
\caption{Same as Figure \ref{fg:samfracham} but using HAM based on
  group-integrated stellar mass, which removes the low cold baryon
  fraction spur.}

\label{fg:samfracham2}
\end{figure*}

In Figure \ref{fg:samfracham} we show stellar and cold baryon
fractions as a function of $M_{halo}^{HAM}$ for the SAM FOF mock
catalog. Using HAM results in the familiar shape seen in the ECO and
RESOLVE-B data in Figure \ref{fg:bfrac}. The baryon fractions peak at
a similar group halo mass, although the behavior at low group halo
mass appears somewhat flatter and more scattered than observed in the
data. The spur seen at high group halo mass and low stellar or cold
baryon fraction is due to a population of galaxies in the SAM with low
mass given their brightness, probably due to inadequate consideration
of dust. Examination of this galaxy population reveals that they are
massive, blue, and star-forming galaxies. To produce observed
magnitudes, the SAM performs stellar population synthesis modeling
based on the star formation and metallicity history of the modeled
galaxy, and then applies a physical model for dust as described in
\citet{2016MNRAS.462.3854L} to produce observed galaxy magnitudes.
The dust fraction, extinction curve shape, albedo, and thus optical
depth are all set to the locally measured value of the Solar
neighborhood, which may not be applicable to galaxies of all
masses. The attenuation of starlight by the dust is computed at all
wavelengths and then redistributed as a blackbody towards infrared
wavelengths. Geometric effects are taken into account by assuming
random orientations of the galaxies. The assumptions used and the
inherent uncertainties when modeling dust may result in
under-extinction of massive, blue, star-forming galaxies. Recent
work by Gonzalez-Perez et al.\ (submitted) also shows that these
massive blue galaxies are too large, and thus the dust density is too
small leading to inefficient absorption of blue and UV light. The
under-attenuation of light in massive galaxies is also apparent in
Figure 30 of \citet{2016MNRAS.462.3854L}, which examines the $g-r$
color distribution of SAM galaxies as a function or brightness
compared to SDSS data, finding that in the most massive bin, the SAM
shows a bimodal color distribution, where the data show only red
galaxies. Moreover, we find that performing HAM with group $M_{star}$
(Figure \ref{fg:samfracham2}) causes the spur to go away, as the
stellar population effects are removed. Group halo masses are then
reduced (typically by $\sim$1 dex) on the x-axis, which also causes
the stellar and cold baryonic fractions to increase on the y-axis.

\begin{figure*}
\plottwo{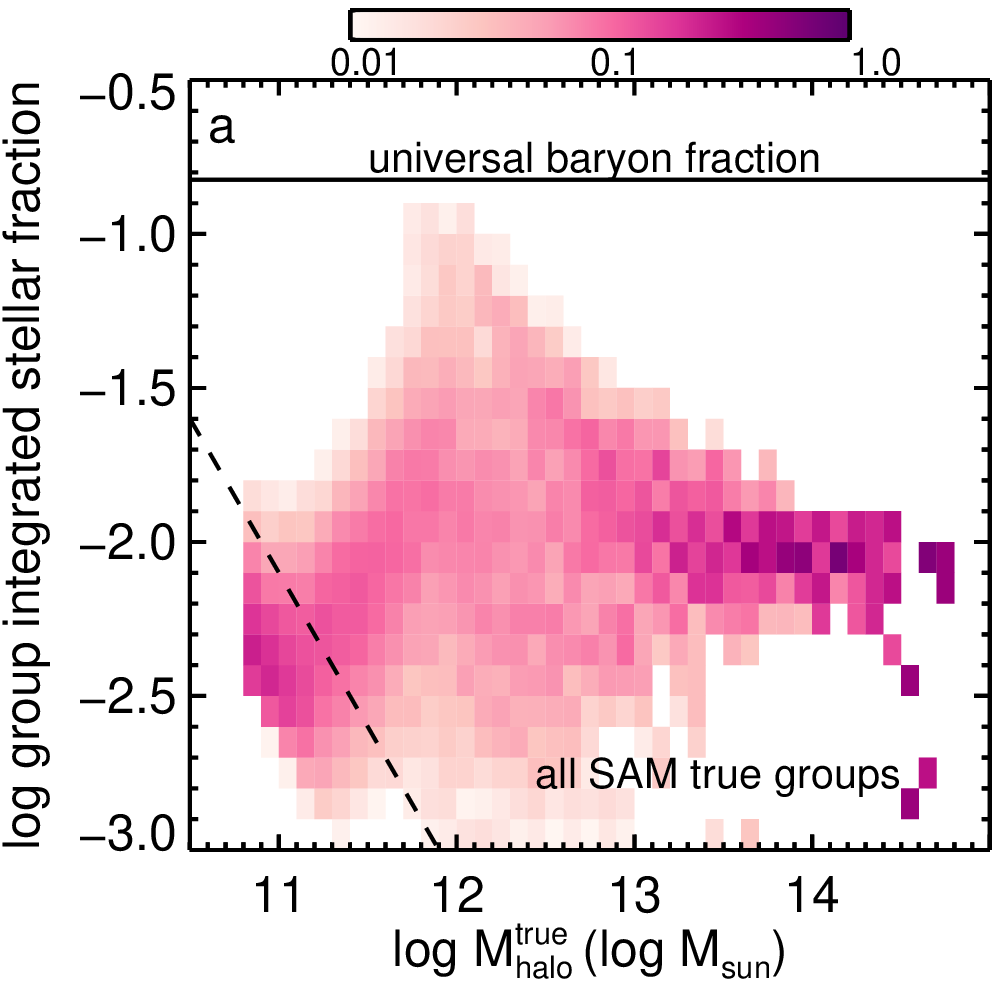}{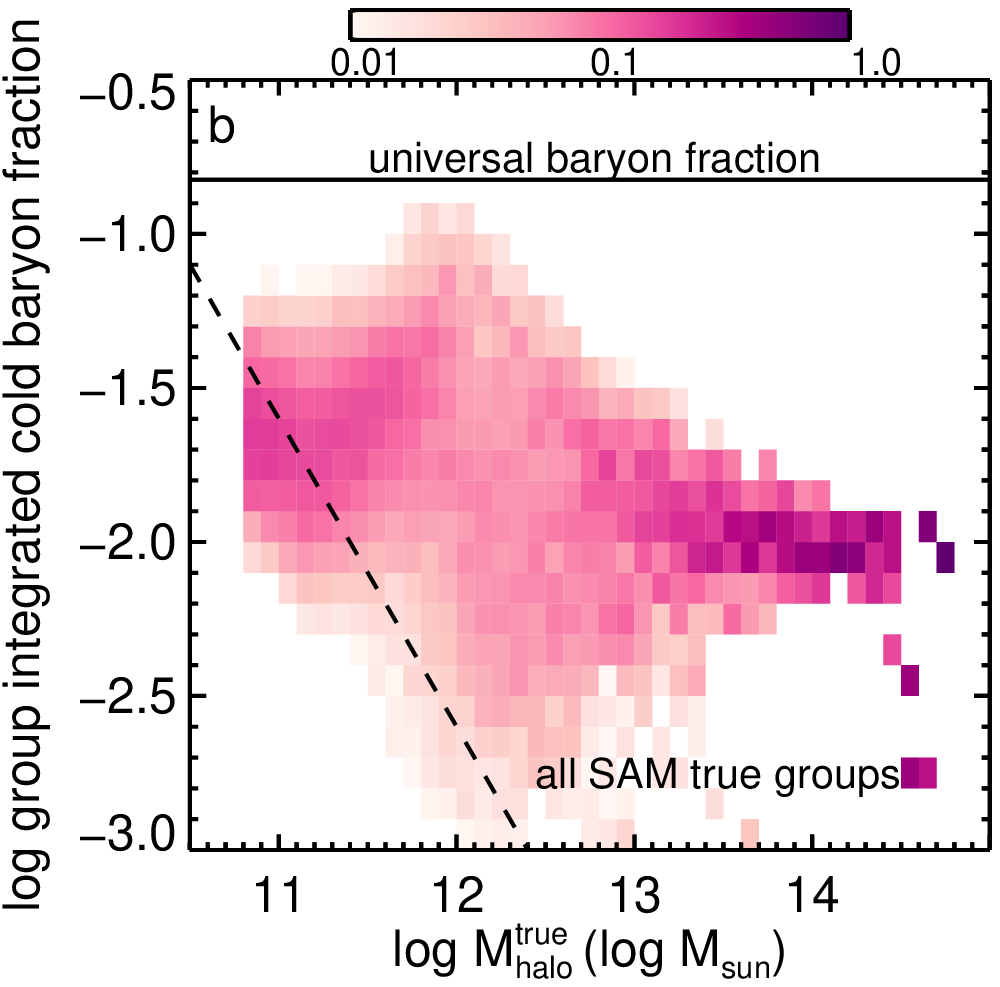}
\epsscale{1.0}
\caption{Conditional density plot of group-integrated (a) stellar and
  (b) cold baryon fractions for the SAM using the true group halo
  mass. We find that below $\sim$10$^{13}$~\msun, the stellar and cold
  baryon fractions become extremely varied, even more so than seen
  with the dynamical mass estimates for ECO and RESOLVE-B.}

\label{fg:samfractrue}
\end{figure*}

Using the true group halo masses (Figure \ref{fg:samfractrue}) for the
SAM yields completely different results. The SAM cold baryon fractions
vary strongly below a halo mass of $\sim$10$^{13}$~\msun{} and reach a
peak dispersion of over 2 dex at \mbox{$M_{halo}$ $\sim$ 10$^{12}$
  \msun}. This dispersion is much larger than what we find in the data
even using the dynamical masses. The SAM suggests that there should be
a population of extremely low cold baryon fraction groups over
precisely the same regime where our data suggests groups are reaching
a maximum baryonic collapse efficiency.

Analysis of the SAM confirms the serious issue with studying cold
baryon fractions of groups using HAM, already suggested by our
dynamical mass analysis: the built-in relationship between group
luminosity (or stellar mass) and group halo mass \textit{produces} the
tight relation between group baryon fraction and group halo
mass. There is evidence that groups can have widely varying ratios of
hot X-ray gas to collapsed baryons \citep{2016MNRAS.455.3628R}, and
the HAM algorithm does not account for this diversity, treating all
groups of similar luminosity (i.e., similar collapsed baryon content)
to be the same. Even with the widely varying cold baryon fractions in
the SAM, using FOF and HAM produces the familiar upside-down U shape
seen in all HAM analyses.

\begin{figure*}
\plottwo{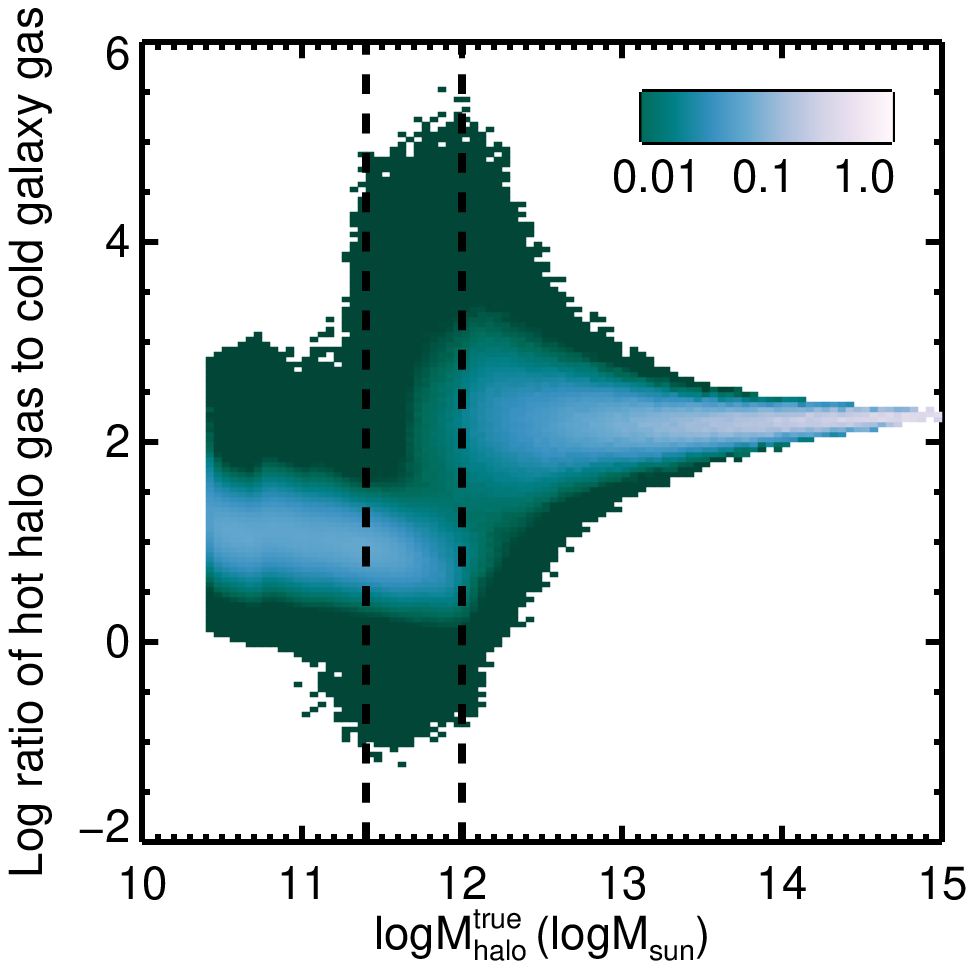}{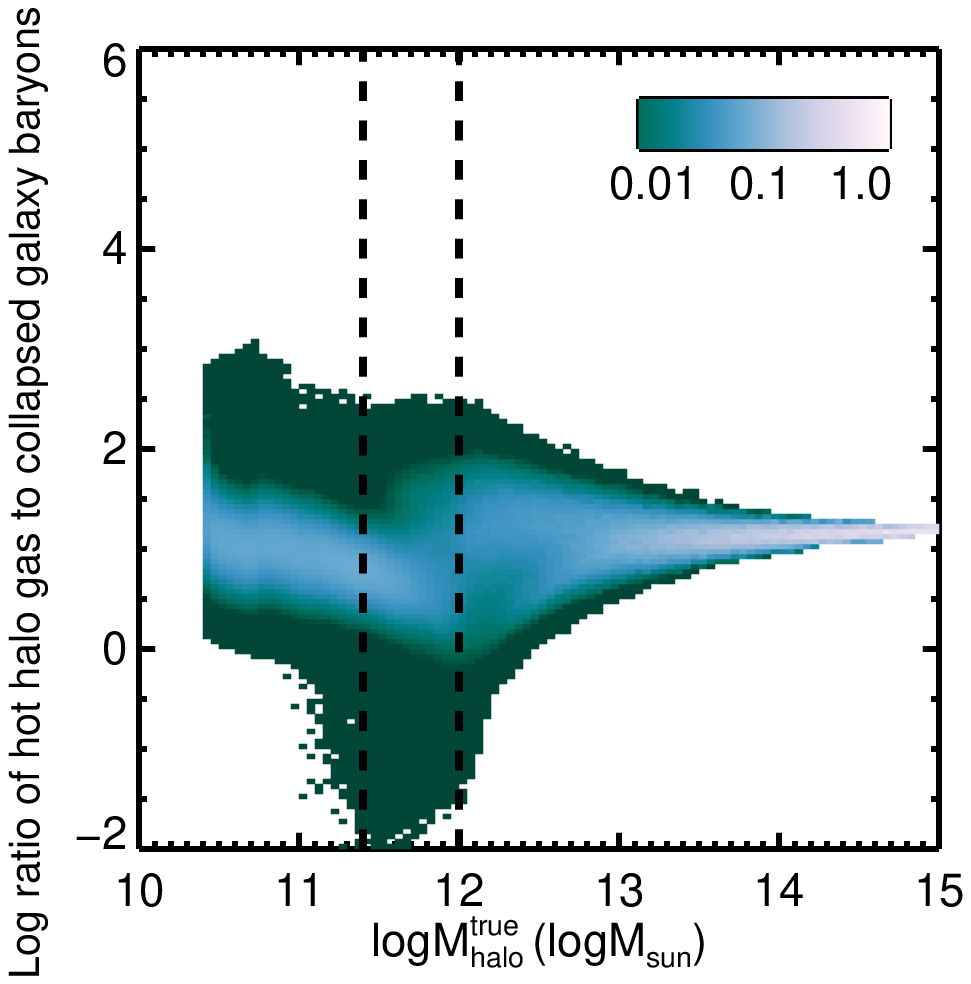}
\epsscale{1.0}
\caption{Conditional density plot of the ratio of hot halo gas to (a)
  cold (collapsed) gas mass and (b) cold (collapsed) baryonic mass in
  galaxies for the SAM true groups. Over the nascent group halo mass
  regime, the scatter in ratio between the hot halo gas to collapsed
  galaxy gas increases as galaxies transition from single objects
  within their halo to larger groups. The overall change in
  hot-to-cold gas fraction between low- and high-mass groups is driven
  by their differing stellar content, as revealed in panel (b), which
  includes the cold gas and stellar content of groups in the
  denominator.}

\label{fg:samhotgasfrac}
\end{figure*}

To highlight this issue, we show the ratio of hot halo gas to cold
(collapsed) gas in galaxies from the SAM mock true groups in Figure
\ref{fg:samhotgasfrac}a. The SAM hot halo gas includes all gas outside
of the galaxy, both halo gas that can be accreted onto galaxies and
gas ejected from the galaxy by feedback. The plot shows that low-mass
groups and high-mass groups have ratios of hot-to-collapsed gas around
$\sim$10 and $\sim$100 respectively, but in the nascent group regime
($\sim$10$^{11.4-12}$~\msun) the ratio of hot-to-collapsed gas becomes
widely varying. It may be that dynamical mass estimates better recover
this scatter in hot-to-cold baryon fractions for lower-mass groups. We
further discuss the wide variation in hot-to-collapsed gas in the SAM
and its relation to the implementation of feedback in the SAM in \S
\ref{sec:nascentgroups}.


Figure \ref{fg:samhotgasfrac}a also shows that low-mass true groups in
the SAM all have ratios of hot-to-collapsed gas mass that are much
lower than in high mass halos. At the same time, the low-mass observed
groups in ECO have low cold baryon fractions using dynamical masses
(Figure \ref{fg:bfracdyn}), which decrease toward lower halo
masses. Taking the theoretical and observational results together may
seem inconsistent with a constant baryon fraction. Figure
\ref{fg:samfractrue}b does appear to be consistent with Figure
\ref{fg:bfracdyn}b in observed and theoretical cold baryon fractions
for low-mass systems. The large change in hot-to-cold gas ratios in
Figure \ref{fg:samhotgasfrac}a is put into perspective by Figure
\ref{fg:samhotgasfrac}b, which shows that much of the change is
driven by the stellar content of low and high mass
groups. Furthermore, the SAM's definition of cold gas, computed by
determining the amount of cooling gas that has had enough time to
reach the center of the halo (thus governed by the cooling and
free-fall times), may overestimate the cold gas because it does not
include the effect of ionization by internal radiation from star
formation within the galaxy. Thus a small amount of the cold gas in
the SAM could actually be ionized gas within the galaxy disk that
would not be observable in HI.

\subsection{Adding The Hot Gas}
\label{sec:discussaddhotgas}

In large groups, the dominant baryonic component is the hot X-ray
emitting gas (e.g.,
\citealp{1977MNRAS.181P..25M,1993Natur.366..429W,2009ApJ...703..982G}),
thus the large offset between the group CBMF and group HMF
at large group halo masses is not unexpected.

To produce a group BMF that includes hot gas mass, we use the hot gas
fraction scaling relation given by \citet{2009ApJ...703..982G}, which
is calibrated on groups and clusters from the COSMOS survey:

\begin{equation}
 f_{gas} = 9.3(\frac{M_{500c}}{2\times10^{14}})^{0.21}
\label{eq:fgas1}
\end{equation}

\noindent
Their calibration, however, was performed for groups with halo masses
defined at $M_{500c}$ (i.e., an overdensity of 500 times the critical
density of the universe). To use their calibration, we must scale the
gas fraction for halo masses defined at $M_{280b}$. We assume that the
dark matter halo mass density follows an NFW profile (equation
\ref{eq:nfw}) and that the hot gas density distribution follows a
beta-model profile (equation \ref{eq:gas}).

\begin{equation}
\rho_{dm} = \frac{\rho_{0,dm}}{\frac{r}{r_s}(1+(\frac{r}{r_s})^2)}
\label{eq:nfw}
\end{equation}

\begin{equation}
\rho_{g} = \rho_{0,g}(r^2 + r_c^2)^{\frac{-3\beta}{2}}
\label{eq:gas}
\end{equation}

\noindent
The scale radius of the NFW profile, $r_s$, is determined by the
virial radius and halo concentration and $\rho_{0,dm}$ is the virial
overdensity of dark matter. For the hot halo gas profile, $r_c$ is the
core radius of the profile, $\rho_{0,g}$ is the gas density
normalization, and $\beta$ is the power law slope. Based on $\beta$
model fits from simulations in \citet{1998ApJ...503..569E}, we set
$r_c$ = $r_s/3$ and $\beta$ = 2/3. We integrate these two density
distributions out to $r_{500c}$, and we use the value of $f_{gas}$ at
$r_{500c}$ (from \citealp{2009ApJ...703..982G}) to determine the ratio
of $\rho_{0,gas}$ to $\rho_{0,nfw}$. Finally we integrate the two
density distributions out to $r_{280b}$ and determine the new
$f_{gas}$ calibration given in equation \ref{eq:fgas2}.

\begin{equation}
f_{gas} = 10.3(\frac{M_{280b}}{2\times10^{14}})^{0.21}
\label{eq:fgas2}
\end{equation}

\begin{figure}
\plotone{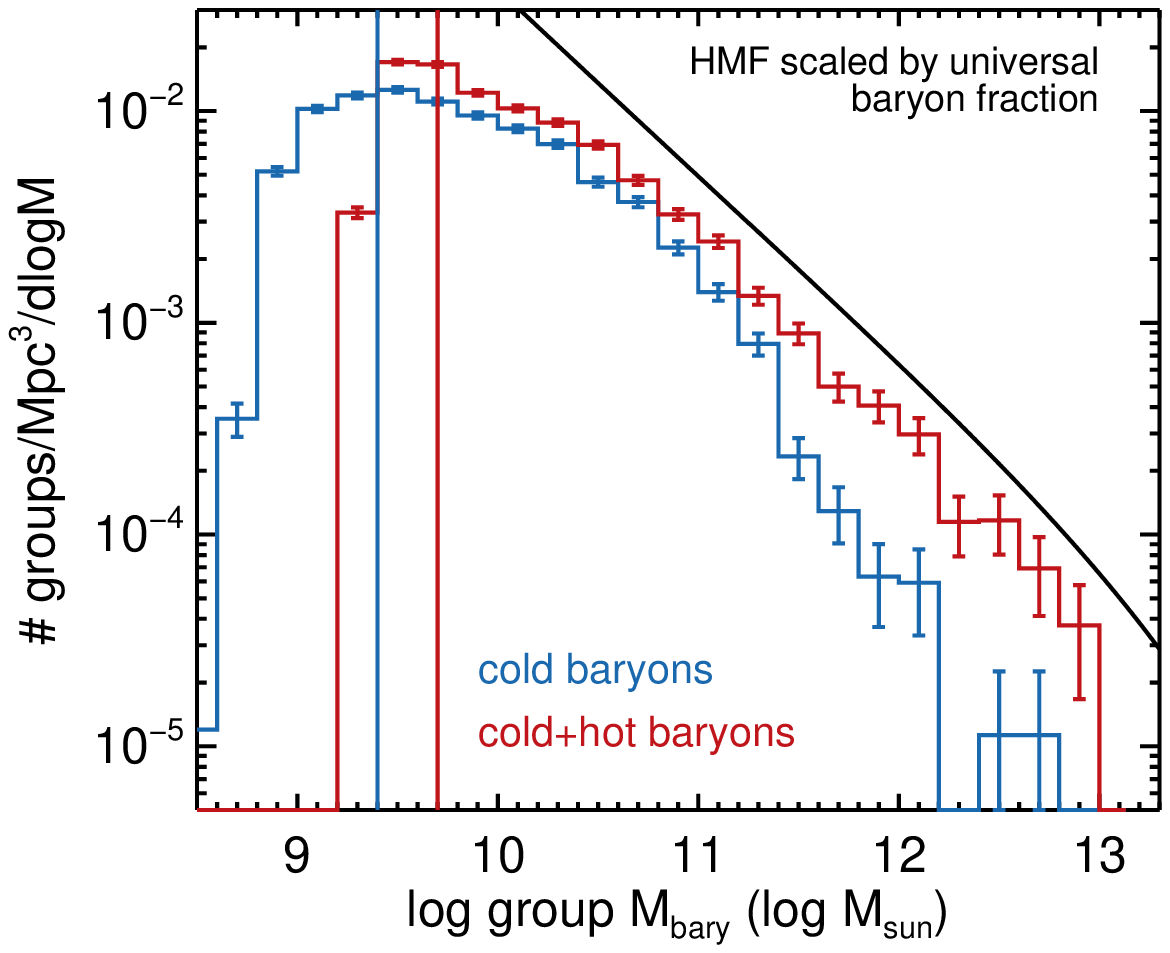}
\epsscale{1.0}
\caption{ECO group CBMF (blue) and BMF including hot X-ray gas (red),
  where we used the hot gas prescription based on dark matter halo
  mass from \citet{2009ApJ...703..982G}, scaled for our halo mass
  definition. The completeness limits are shown as vertical lines
  corresponding to the colors of the histograms. The BMF completeness
  limit is determined by finding the maximum hot gas correction to
  group $M_{bary}^{cold}$ at 10$^{9.4}$~\msun, which we find to be
  $\sim$0.3 dex. Including the hot gas causes the group BMF to run
  parallel to the dark matter HMF. In this analysis for simplicity we
  do not use full group mass likelihood distributions but instead
  median values. Error bars represent the Poisson statistics.}
\label{fg:bmfphg}
\end{figure}

In Figure \ref{fg:bmfphg}, we show the group CBMF (blue) and BMF
including hot gas (red) for ECO using the HAM mass estimates to
determine the hot gas component. (Results are similar using the
dynamical mass estimates.) We compute both the group CBMF and the
group BMF using the median group cold baryonic mass measurement for
simplicity (using Poisson statistics to compute error bars).
Including the hot gas significantly changes the shape of the group BMF
at high halo masses, causing it to track the HMF down to group
\mbox{$M_{bary}$ $\sim$10$^{11}$~\msun}.

The fact that the group BMF does not line up exactly with the group
HMF (by a factor of $\sim$2) at high masses suggests that there is
still some missing baryonic component that we have not
included. Possibilities include: 1) intracluster light that
contributes $\sim$10\%
\citep{2004ApJ...609..617F,2006AJ....131..168K}, 2) low-mass satellite
galaxies below our mass limits (contributing at most 14\%, see \S
\ref{sec:grpint}), and 3) warm-hot gas, WHIM, too cool to emit in
X-rays, which potentially contributes 40\%--50\% of baryons based on
simulations \citep{2001ApJ...552..473D,2006ApJ...650..560C}. We also
note that most studies of the hot X-ray gas in clusters find that
accounting for the hot X-ray gas and stars does not completely account
for all the baryons
\citep{2007ApJ...666..147G,2009ApJ...703..982G,2015arXiv151007046M}.

Another consideration is that much of the WHIM may lie outside the
viral radius of the halo due to supernova-driven outflows. From
simulations of Milky Way galaxies, \citet{2016ApJ...819...21S} find
that 20\%--30\% of the WHIM is pushed out between 1--3 virial radii, and
that 90\% of the universal baryon fraction is recovered only when
considering the halo gas out to 3 virial radii (much further than the
considerations used in our analysis). Examining simulations of lower
mass galaxies (\mbox{$M_{halo}$ $\sim$ 10$^{11}$ \msun}),
\citet{2017MNRAS.466.4858W} also find that baryons are expelled beyond
twice the virial radius. Such feedback may also cause groups groups to
begin forming with a depleted baryon fraction
\citep{2016MNRAS.456.4266L}. Thus the expectation that we should be
able to account for all the baryons in groups and clusters to the
virial radius may be false, and the shortfall, while interesting, may
not necessarily be problematic.

\section{Discussion}
\label{sec:discussion}

In this section, we present further discussion of the results of this
work. First, we consider the relationship between baryonic collapse
efficiency and galaxy and group growth, examining in particular the
nascent group regime. Second, we discuss possible undetected baryons
in galaxies and their halos and consider the effects of these mass
components on the group BMF.


\subsection{Baryonic Collapse Efficiency and Galaxy Growth: From Isolated Dwarfs to Nascent Groups}
\label{sec:nascentgroups}

At the lowest group halo masses ($<$$10^{11.4}$ \msun), we have seen
that groups are mostly isolated dwarf galaxies in \mbox{$N = 1$} halos
and are \textit{increasing their cold baryonic mass faster than their
  dark matter halo mass} (Figures \ref{fg:bfrac} to
\ref{fg:bfracdyn}). While isolated dwarfs are often thought of as
inefficient star formers (due to their large HI reservoirs, which lead
to long gas depletion times), K13 showed that they are nonetheless
growing rapidly using a long-term measure of galaxy growth called
fractional stellar mass growth rate (FSMGR), defined as the stellar
mass produced in the last Gyr divided by the stellar mass produced
prior to that Gyr. Isolated dwarf galaxies have FSMGR $\sim$ 1,
implying stellar mass doubling on Gyr timescales. Additionally,
\citet{2013MNRAS.428.3121M} show that isolated dwarf galaxies in
low-mass halos are growing much more rapidly than their halos at
current times using multi-epoch abundance matching. Thus, while not at
the peak of cold baryon fraction, such low mass groups are at peak
galaxy growth rates.

Over the halo mass regime of 10$^{11.4-12}$~\msun, the nascent group
regime, we find that groups reach peak collapsed baryon fraction or
``baryonic collapse efficiency.'' Here we use ``efficiency'' in the
usual convention as a level reached rather than a rate of processing
(e.g., \citealp{2009ApJ...696..620C,2012apj...744..159L}). As baryonic
collapse efficiency peaks, galaxy formation slows and group processes
begin to shape the population. Indeed, K13 found that central galaxies
in the nascent group regime (using the central galaxy mass -- group
mass relationship), had lower FSMGR than isolated dwarfs, implying
slowed galaxy growth at these group mass scales. Additionally, from
examination of the galaxy mass functions in E16, we find that nascent
groups may already experience merging and/or stripping of satellites,
as the satellite mass function is depressed relative to the central
mass function and has a flat low-mass slope (to our completeness
limits). Once groups reach this nascent group scale, group processes
such as merging and stripping seem to act to stop the growth of the
collapsed baryonic content of groups. Group cold baryon fractions then
drop towards higher group halo masses, as the uncollapsed hot halo gas
dominates the baryonic content of the halo. Future planned studies of
FSMGR as a group-integrated quantity and as a function of group mass
may help shed more light on the connection between galaxy growth and
group formation.

Connecting these results with the group CBMF (Figure \ref{fg:smfbmf}),
we find that low-mass groups (with peak galaxy formation rates)
have group \mbox{$M_{bary}^{cold}$ $<$ 10$^{10}$ \msun} (from Figure
\ref{fg:mbvmg}a), and thus lie on the shallow power-law slope of the
group CBMF.  Nascent groups have group $M_{bary}^{cold}$ ranging from
$\sim$10$^{10-10.8}$~\msun, which places them just below the knee of
the group CBMF. The highest mass halos then lie on the steep falloff
towards higher masses (although accounting for their hot gas the
falloff has a slope similar to the HMF; Figure \ref{fg:bmfphg}).  That
the nascent groups exist right at this change in mass function
behavior reflects the fundamental change in baryonic collapse
efficiency between low- and high-mass group halos.

We have shown that HAM halo mass estimates yield a tight maximum in
baryonic collapse efficiency, but dynamical halo mass estimates
suggest more scatter. The scatter in collapsed baryon fraction using
dynamical masses is still much smaller than suggested by the SAM true
groups. The discrepancy may be partially due to the reliance on HAM
for $N = 1$ and $2$ groups, for which we have no other proxy for the
halo mass. The SAM, in addition, may have overly large feedback (as
discussed below) causing unrealistic scatter in cold baryon fractions.
To fully discriminate between these two possibilities, we require
independent measurements of the halo masses of $N = 1$ and $2$
groups. The RESOLVE survey is conducting a census of velocity
measurements (either resolved rotation curves or velocity dispersions,
depending on galaxy type), which can serve as a proxy for $N = 1$
group halo masses. Using these measurements, we will in the future
attempt to determine how large the scatter in cold baryon fraction is
at low masses and whether it can be attributed to growing diversity in
the ratio of hot halo gas to cold collapsed baryons.

In the SAM, the transition from isolated dwarfs to larger groups over
the nascent group regime is strongly governed by the model's
implementation of feedback. The nascent group regime is where the
dominant mode of feedback transitions from stellar feedback (at lower
halo masses) to AGN feedback (at higher halo masses). The location of
the knee at higher masses in the SAM (and resulting underprediction of
groups in the SAM near the knee of ECO's group SMF and CBMF, Figure
\ref{fg:samsmfbmf}) reveals that the model's implementation of
feedback is overly efficient at these scales. Recently,
\citet{2016MNRAS.456.1459M} compared the \textsc{galform} galaxy SMF
with observations, finding that the location and amplitude of the knee
of the galaxy SMF are very sensitive to the feedback prescription in
the SAM and that reducing the stellar feedback efficiency improves
agreement in the knee with observed galaxy SMFs.

Additionally, the SAM true groups exhibit extremely varied cold baryon
fractions and hot-to-collapsed gas ratios over the nascent group
regime (see Figures \ref{fg:samfractrue} and
\ref{fg:samhotgasfrac}). This scatter could be partially driven by the
transition from stellar to AGN feedback, which occurs over the nascent
group regime as galaxies transition to the group environment. Some of
the variation may also be a result of the stellar feedback
implementation in \textsc{galform}, which ties the fraction of cold
gas ejected from the disk (the mass loading factor) to the circular
velocity of the galaxy disk or bulge for quiescent or starburst
star-formation episodes respectively. Comparing the central galaxy
stellar to halo mass (SHM) relationship from \textsc{galform} and
\textsc{l-galaxies}, a different SAM described in
\citet{2011MNRAS.413..101G} that ties the mass loading factor to the
halo circular velocity, \citet{2016MNRAS.461.3457G} find larger
scatter in the SHM relationship of \textsc{galform} than that of
\textsc{l-galaxies}. Additionally, \citet{2016MNRAS.456.1459M} find
that implementing the \textsc{l-galaxies} mass loading factor
prescription in \textsc{galform} reduces the scatter in the SHM
relationship. Thus, accurately quantifying the scatter in cold baryon
fractions (or the SHM relationship) over nascent group scales can
provide important constraints that lead to improved models of galaxy
formation.



\subsection{Undetected Gas}
\label{sec:undgas}

So far we have only considered the effect of hot X-ray gas that
dominates the mass of large groups and clusters. There are other gas
components, however, that may especially affect galaxy and group mass,
such as opaque HI gas, CO-traced molecular gas, CO-dark molecular gas,
and warm-hot ionized gas (WHIM). We have not taken into account HI
self-absorption (opaque HI), which could contribute up to 30\% of the
HI gas in the most edge-on galaxies \citep{1994AJ....107.2036G}. We
have also neglected the CO-traced molecular gas in this work, under
the assumption that it rarely dominates the cold gas in galaxies. This
assumption is not true for some large spiral galaxies, but those are
generally dominated by their stellar mass (K13;
\citealp{2014A&A...564A..66B}).


In low-mass, low-metallicity, gas-rich galaxies molecular gas is
difficult to detect with standard tracers like CO
\citep{1998AJ....116.2746T,2005ApJ...625..763L,2012AJ....143..138S,2013ARA&A..51..207B},
due to the low dust content, which allows the CO to be
photo-dissociated more easily. The molecular hydrogen on the other
hand, is self-shielded, making it traceable by the [CI] and [CII]
lines \citep{2006A&A...451..917R,2015MNRAS.448.1607G}. Ongoing work
using the Herschel Dwarf Galaxy Survey seeks to uncover CO-dark gas
through examination of these far IR lines, potentially finding between
ten and several hundred times as much CO-dark as CO-traced molecular
hydrogen \citep{2016arXiv160304674M}. While the CO-dark molecular gas
is mainly thought to contribute significantly to the gas mass of
low-metallicity dwarfs, \citet{2013A&A...554A.103P} find that
$\sim$30\% of the molecular gas mass in the Milky Way could be CO-dark
molecular gas. In fact, the theoretical model of
\citet{2010ApJ...716.1191W} predicts a CO-dark gas fraction of 0.3 for
Milky Way like extinction, and an increasing CO-dark gas fraction for
decreasing extinction values, which may be more applicable to
low-metallicity dwarf galaxies.

Another contribution to the undetected gas component is the
WHIM. Based on simulations, WHIM gas in the galaxy halo contributes
significantly to the baryon census ($\sim$40\%--50\%,
\citealp{2006ApJ...650..560C,2011ApJ...731....6S}), although a
significant amount of WHIM may be pushed outside the virial radius in
Milky Way size halos \citep{2016ApJ...819...21S}. Observations of
X-ray absorption lines have yielded large variations in the mass and
distribution of the WHIM in the Milky Way: from
$\sim$4$\times$10$^{8}$~\msun{} within 20~kpc
\citep{2007ApJ...669..990B} to $\sim$10$^{10}$~\msun{} within 100~kpc
\citep{2012ApJ...756L...8G}.



Studies of the baryonic Tully-Fisher relation suggest that there could
be missing gas in the galaxy disk that scales with the HI gas
component
\citep{2005A&A...431..511P,2008MNRAS.386..138B,2009A&A...501..171R}. These
studies find that a multiplicative factor of 3--11 applied to the HI
mass produces a tighter Tully-Fisher relation across dwarf to giant
galaxy scales. Additional indirect evidence for such undetected gas is
also seen in rotation curve decomposition analyses
\citep{2001MNRAS.323..453H,2012MNRAS.425.2299S}, which find a direct
scaling of the HI gas or baryonic distribution can explain the galaxy
rotation curve. While the above-described potential forms for the
undetected gas may not explain such large multiples of the HI mass,
they do all point to a missing reservoir that could contribute
significantly to the galaxy mass.

\begin{figure}
\plotone{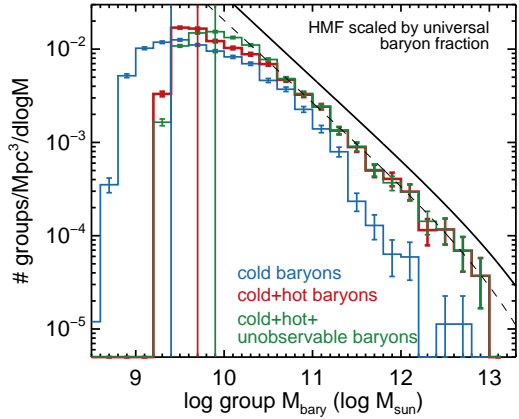}
\epsscale{1.0}
\caption{Group CBMF for ECO (blue), BMF including hot gas (red, based
  on HAM), and BMF including both hot gas and potentially undetected
  gas in low-mass, gas-rich galaxies (green). We find agreement
  between the BMF including hot gas and the HMF rescaled by a 0.07 (a
  factor of 2 smaller than the universal baryon fraction, dashed black
  line) down to $M_{bary}$ = 10$^{10.7}$ \msun. By scaling low-mass,
  gas-rich galaxies, which may harbor large undetected gas components,
  we extend this agreement to $M_{bary}$ = 10$^{10.3}$ \msun.}
\label{fg:bmfphgamg}
\end{figure}

To examine the effect of including such undetected gas in our group
mass functions, we first note that scaling the HMF by a factor of 0.07
(the universal baryon fraction 0.15 divided by 2, dashed black line in
Figure \ref{fg:bmfphgamg}) yields agreement between the HMF and the
group BMF that includes hot gas (red histogram) down to group
\mbox{$M_{bary}$ $\sim$ 10$^{10.7}$ \msun}. This factor of 2 reduction
is in rough agreement with expectations that the WHIM contributes
40\%--50\% of the baryons in group halos. Figure \ref{fg:bmfphgamg} also
shows the effect of including undetected galaxy gas that scales with
the galaxy HI gas component (green histogram). We have only scaled the
cold baryonic masses of low-mass (\mbox{$M_{star}$ $<$ 10$^{9.5}$
  \msun}), gas-rich (\mbox{1.4$M_{HI}$/$M_{star}$ $>$ 1.0}) galaxies,
which are the most likely to harbor a significant amount of undetected
CO-dark gas. We multiply their cold HI gas mass by a factor of 2. Such
scaling affects the low-mass end of the group BMF and improves
agreement with the scaled HMF down to \mbox{$M_{bary}$ $\sim$
  10$^{10.3}$ \msun}. In these analyses, as for our hot gas analysis,
we have also used the median stellar and gas mass (baryonic-stellar)
measurements rather than the full likelihood distribution for
simplicity.


Above 10$^{11}$~\msun, there is not much change in the total group
BMF, since there are few low-mass, gas-rich galaxies in high mass
halos. Below 10$^{11}$~\msun, the group BMF including hot gas and this
scaled cold gas component does continue to track the scaled HMF
better, lending support to the idea that undetected gas may contribute
significantly to the galaxy component of the group mass, particularly
at nascent group scales and below.

\section{Conclusions}
\label{sec:conclusions}
	
In this work, we have examined the group-integrated stellar and
baryonic content of groups for the RESOLVE and ECO surveys. We have
further compared with results from a SAM mock catalog and discussed
implications for group formation particularly in the nascent group
regime.

\textbullet{} The group SMF and CBMF exhibit steep slopes at high
masses, and a shallower rising slope at low masses. They most closely
approach the dark matter HMF near $\sim$10$^{11}$~\msun. The low-mass
slope of the group CBMF is steeper than that of the group SMF, but
still deviates from the steep dark matter HMF slope (see \S
\ref{sec:groupmfs} and Figure \ref{fg:smfbmf}).

\textbullet{} The SAM's group SMF and CBMF are similar to those of ECO
at high masses. However, the SAM has fewer groups at the transition
mass of $\sim$10$^{11}$~\msun, and the SAM mass functions rise more
steeply at low-masses (Figure \ref{fg:samsmfbmf}). These differences
are likely due to the sensitivity to the transition between stellar
and AGN feedback in the models (\S \ref{sec:nascentgroups}).

\textbullet{} Inclusion of hot halo gas in the group BMF using a
literature prescription produces a slope that runs parallel to the
dark matter HMF (although still low by a factor of 2, Figure
\ref{fg:bmfphg}). The hot halo gas does not contribute significantly
to the group baryonic mass below $\sim$10$^{11}$~\msun{} (see \S
\ref{sec:discussaddhotgas}).

\textbullet{} If we assume that there is additional undetectable gas
in galaxies that scales with the HI mass for low-mass, gas-rich
galaxies (which may have large reservoirs of CO-dark molecular gas),
adopting a multiplicative factor of 2 as has been suggested by
baryonic Tully-Fisher studies, we can produce a more steeply rising
low-mass slope below 10$^{11}$~\msun{} that continues to run parallel
to the dark matter HMF (\S \ref{sec:undgas} and Figure
\ref{fg:bmfphgamg}).

\textbullet{} Examination of the stellar and cold-baryon fractions as
a function of HAM group halo mass reveals the familiar upside-down U
shape seen in previous work (Figures \ref{fg:bfrac} and
\ref{fg:bfrac2}). While the group stellar fraction has a narrow peak
near $\sim$10$^{11.8}$ \msun, the peak of the group cold-baryon
fraction (or baryonic collapse efficiency) is spread across a broader
halo mass range of $\sim$10$^{11.4-12}$~\msun{}. This broad peak of
baryonic collapse efficiency coincides with the nascent group regime,
wherein galaxies and galaxy groups transition from gas-rich to stellar
dominated.

\textbullet{} Because HAM halo masses enforce monotonicity between
group halo mass and collapsed baryons as quantified by group $L_r$ or
M$_{star}$, but we are interested in diversity of collapsed baryon to
halo mass ratios, we have developed a new way of measuring dynamical
masses that allows us to probe halo mass independent of collapsed
baryon properties. Our hybrid halo masses smoothly transition from
using HAM for $N = 1$ and $2$ groups, stacked dynamical
masses for $N > 2$ groups, and individual group dynamical masses for
the highest $N$ groups.

\textbullet{} Examination of the stellar and cold-baryon fractions as
a function of dynamical halo mass estimates suggests more scatter in
collapsed baryon fractions than observed using HAM halo mass estimates
(Figure \ref{fg:bfracdyn}), potentially reflecting variations in the
hot-to-collapsed baryon fraction between groups at fixed group
mass. This result argues for caution in interpreting baryon fractions
using HAM, as the built-in assumption of mass following collapsed
baryons may break down across intermediate group halo mass regimes.

\textbullet{} The SAM true groups also suggest that there should be a
population of very low cold baryon fraction groups (Figure
\ref{fg:samfractrue}). Once HAM is performed on the SAM, we obtain the
same upside down U shape as seen in the data (Figures
\ref{fg:samfracham} and \ref{fg:samfracham2}). This result underscores
the importance of recognizing the built-in relationship between halo
mass and group cold baryonic mass when using HAM halo mass estimates.

The results from this paper touch on several aspects of the baryon
census. For example, we have shown that the group BMF can obtain a
similar shape as the HMF once the collapsed baryonic matter within
groups is combined with the hot halo gas and potential CO-dark gas in
gas-rich, dwarf galaxies. Although the group BMF is shifted lower in
mass by a factor of $\sim$2, that shift is in agreement with WHIM
estimates of 40\%--50\%, and this result suggests that at the scales we
probe ($M_{halo}$ $\sim$ 10$^{11-14.5}$ \msun), we can account for
most of the baryons. Using dynamical masses to explore group cold
baryon fractions, however, points to far more variation in the
hot-to-collapsed baryon ratio in groups than implied by using HAM,
especially across the nascent group regime. The SAM provides
additional support for large variations in hot-to-collapsed baryon
fractions at nascent group scales. Nascent groups appear to be sites
of active group formation processes such as merging and stripping, as
shown by the depressed, flat low-mass slope of the nascent group
galaxy mass function in E16. The nascent group regime, however, is
where our stacked dynamical mass analysis starts to break down as we
approach the acutely low-$N$ regime, thus leaving us unable to fully
probe these variations. In future work, we plan to measure dynamical
masses using internal galaxy kinematics to extend our analysis to the
lowest-$N$ groups, enabling combined analysis of the galaxy and group
(subhalo and halo) velocity functions.



\acknowledgements

We thank Matt Bayliss for giving us the idea of stacking groups
to estimate dynamical masses at low $N$. We also thank Brian
McNamara for useful discussions of group baryon fractions. We would
like to thank Prajwal Kafle and Laura Parker for helpful conversations
regarding group dynamical mass estimation and virialization status. We
are grateful to Maud Galametz and Aleksandra Sokolowska for
discussions of CO-dark molecular gas in galaxies and ionized gas in
halos respectively. We also thank Peter Mitchell for discussions of
feedback related to the knee of the group mass function. K.D.E.\ and
S.J.K.\ acknowledge support for this research from NSF grant
AST-0955368. K.D.E also acknowledges support from NSF grant
OCI-1156614, the North Carolina Space Grant Fellowship, the University
of North Carolina Royster Society of Fellows, and Sigma Xi
Grants-in-Aid of Research Program. This work is based on observations
from the SDSS. Funding for SDSS-III has been provided by the Alfred
P. Sloan Foundation, the Participating Institutions, the National
Science Foundation, and the U.S. Department of Energy Office of
Science. The SDSS-III web site is http://www.sdss3.org/. SDSS-III is
managed by the Astrophysical Research Consortium for the Participating
Institutions of the SDSS-III Collaboration including the University of
Arizona, the Brazilian Participation Group, Brookhaven National
Laboratory, Carnegie Mellon University, University of Florida, the
French Participation Group, the German Participation Group, Harvard
University, the Instituto de Astrofisica de Canarias, the Michigan
State/Notre Dame/JINA Participation Group, Johns Hopkins University,
Lawrence Berkeley National Laboratory, Max Planck Institute for
Astrophysics, Max Planck Institute for Extraterrestrial Physics, New
Mexico State University, New York University, Ohio State University,
Pennsylvania State University, University of Portsmouth, Princeton
University, the Spanish Participation Group, University of Tokyo,
University of Utah, Vanderbilt University, University of Virginia,
University of Washington, and Yale University. This work is based on
observations made with the NASA Galaxy Evolution Explorer. GALEX is
operated for NASA by the California Institute of Technology under NASA
contract NAS5-98034.  This publication makes use of data products from
the Two Micron All Sky Survey, which is a joint project of the
University of Massachusetts and the Infrared Processing and Analysis
Center/California Institute of Technology, funded by the National
Aeronautics and Space Administration and the National Science
Foundation. This work is based in part on data obtained as part of the
UKIRT Infrared Deep Sky Survey.  This work uses data from the Arecibo
observatory. The Arecibo Observatory is operated by SRI International
under a cooperative agreement with the National Science Foundation
(AST-1100968), and in alliance with Ana G. M\'endez-Universidad
Metropolitana, and the Universities Space Research Association. This
work is based on observations using the Green Bank Telescope. The
National Radio Astronomy Observatory is a facility of the National
Science Foundation operated under cooperative agreement by Associated
Universities, Inc.


\appendix
\label{app:app1}
In this appendix, we describe the procedure used to identify and break
up false pairs of galaxies in the ECO, RESOLVE-B, and SAM FOF group
catalogs. We use a mock catalog built on a $\Lambda$CDM N-body
simulation, in which central and satellite galaxies have been placed
into halos according to the HOD framework of
\citet{2002ApJ...575..587B} and redshift space distortions have been
included (see M15 for more details).

With this mock catalog, we know the true identities of group halos,
and we run FOF group finding to examine how the algorithm affects
group assignments. Comparing the true and FOF groups, we find that the
pair population is often affected by group finding errors. A large
portion of FOF pairs are ``false pairs,'' or two $N = 1$ galaxies
merged together. We additionally find ``split-off'' FOF pairs that
were originally two members of a larger true group and were split off
in the FOF group finding process.  Matching between the true and FOF
groups, we identify among FOF groups true pairs, false pairs, and
split-off pairs, finding that they comprise 62\%, 22\%, and 16\% of
the FOF pair population respectively.

\begin{figure}
\epsscale{0.7}
\plotone{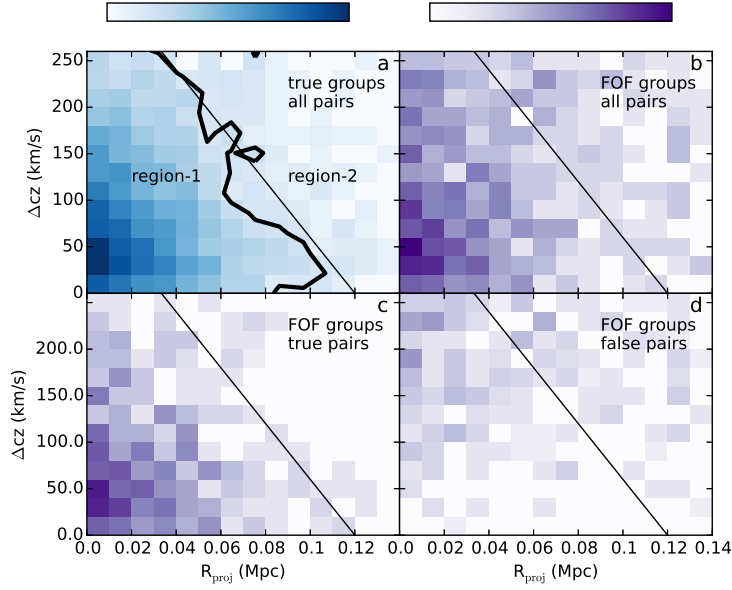}

\caption{Mock catalog pair statistics, shown as 2-D histograms of the
  distribution of pairs ($N = 2$ groups) over $\Delta$cz
  vs.\ $R_{proj}$ space for (a) all true pairs, (b) all FOF pairs, (c)
  true FOF pairs, and (d) false FOF pairs. (Split-off FOF pairs are
  not shown, but are distributed roughly equally over this space.) The
  2-D histograms for panels (b)--(d) are normalized to all FOF pairs,
  with the color bar shown above panel (b). In panel (a) we show the
  contour containing 95\% of the true-mock pairs (thick black line) as
  well as the line we choose to define region-1 and region-2 (thin
  black line). Only 5\% of true FOF pairs live in region-2, while
  nearly 50\% of false FOF pairs live in region-2. We therefore break
  up all pairs in region-2, improving the percentage of true pairs
  among FOF pairs from 62\% to 73\%.}
\label{fg:app}
\end{figure}

To break up the false pairs, we examine the distribution of true pairs
in $\Delta$cz--$R_{proj}$ space, where $\Delta$cz is the difference in
redshift between the pair galaxies and $R_{proj}$ is computed for the
pair as described in \S \ref{sec:dynhn}. The 2-D histogram of true
pairs is shown in Figure \ref{fg:app}a. We then find the contour that
encloses $\sim$95\% of the true pairs (bold black line). To
simplify our region, we draw a line reflecting that contour and
designate ``region-1, '' which contains 95\% of true pairs, and
``region-2,'' which contains the remaining 5\% of true pairs.

In panels (b)--(d) of Figure \ref{fg:app}, we show the 2-D histograms
of overall, true, and false pairs among the FOF groups (normalized to
the overall FOF pair population). (Split-off pairs are distributed
roughly evenly over the $\Delta$cz--$R_{proj}$ space.) In the overall
FOF pair population, 5\%, 48\%, and 40\% of true, false, and split-off
pairs reside within region-2. Based on this analysis, we split up all
FOF pairs residing in region-2, thereby removing $\sim$1/2 of the
false pairs at the expense of splitting up 5\% of the true
pairs. (Split-off pairs in region-2 also now contribute to the FOF $N
= 1$ population, however, they make up $<$2\% of the FOF $N = 1$
population.) Our new FOF pair population now consists of 73\% true
pairs, 15\% false pairs, and 12\% split-off pairs.

\end{document}